\documentstyle[multicol,prb,aps,epsfig]{revtex}

\begin{document}

\title{Polymer induced depletion  potentials in polymer-colloid
mixtures }

\author {A.A. Louis$^1$, P.G. Bolhuis$^2$, E.J. Meijer$^2$ and
J.P. Hansen$^1$,} \address{$^1$Department of Chemistry, Lensfield Rd,
Cambridge CB2 1EW, UK} \address{$^2$Department of Chemical
Engineering, University of Amsterdam, Nieuwe Achtergracht 166, NL-1018
WV Amsterdam, Netherlands.}  \date{\today} \maketitle
\begin{abstract}
\noindent The depletion interactions between two colloidal plates or
between two colloidal spheres, induced by interacting polymers in a
good solvent, are calculated theoretically and by computer
simulations.  A simple analytical theory is shown to be quantitatively
accurate for case of two plates.  A related depletion potential is
derived for two spheres; it also agrees very well with direct computer
simulations.
Theories based on ideal polymers show important deviations with
increasing polymer concentration: They overestimate the range of the
depletion potential between two plates or two spheres at all densities,
with the largest relative change occurring in the dilute regime.  They
underestimate the well depth at contact for the case of two plates,
but overestimate it for two spheres.  Depletion potentials are also
calculated using a coarse graining approach which represents the
polymers as ``soft colloids'': good agreement is found in the dilute
regime.  Finally, the effect of the polymers on colloid-colloid
osmotic virial coefficients is related to phase behavior of
polymer-colloid mixtures.
\end{abstract}
\begin{multicols}{2}

\section{Introduction}
Effective potentials are a key to unlocking the equilibrium
 behavior of many soft-matter systems\cite{Liko01,Loui01a}.
The basic philosophy behind this coarse-graining approach is that the
initial effort in deriving these potentials is recouped when they are
input into the well oiled machinery of liquid state
theory\cite{Hans86}, or when they are used in computer
simulations\cite{frenkelbook}.  An archetypal example is the
depletion potential, induced between colloidal particles by
non-adsorbing polymers.  Asakura and Oosawa\cite{Asak54} first showed
that a bath of such polymers, characterized by their radius of
gyration $R_g$, induces an attractive depletion interaction of range
$D\approx 2 R_g$ between two plates.  Their calculation was exact
for non-interacting polymers.  Later, the same authors, and
independently Vrij\cite{Asak58}, derived a depletion potential between
two colloidal hard spheres (HS) by approximating the (ideal) polymers
as penetrable spheres.  This is often termed the Asakura-Oosawa (AO)
model.

A good example of the effective potential coarse-graining approach is
the calculation of the phase behavior of polymer-colloid mixtures by
Gast, Hall, and Russel\cite{Gast83}, and also by Lekkerkerker {\em et
al.}\cite{Lekk92} and Meijer and Frenkel\cite{Meij94}.  They found,
using an AO depletion potential approach, that the fluid-fluid phase
line of colloids of radius $R_c$ becomes metastable w.r.t.\ the
fluid-solid phase line if the size ratio $q=R_g/R_c$ is less than
about $0.35$, in qualitative agreement with
experiments\cite{Cald93,Ilet95}.  Their work demonstrates how an
accurate knowledge of the depletion potential can lead to a good
understanding of the equilibrium behavior of colloid polymer mixtures.
The latter are important not only because of their relevance to many
industrial and biological processes, but also because they form an
important model system for equilibrium and non-equilibrium behavior in
soft matter science.

Whereas the depletion interaction for ideal polymers is now
quantitatively understood, the depletion interaction induced by
polymers with excluded volume interactions is only qualitatively
understood.  Experiments on the phase behavior\cite{Cald93,Ilet95} and
structure\cite{Weis99,Mous99} of polymer colloid mixtures also show
deviations from the simple AO model\cite{Loui99a,Dijk99}, as do direct
measurements of the depletion potentials\cite{Ohsh97,Verm98,Bech99}.
Theoretical attempts to directly calculate the depletion potentials
for interacting polymers include scaling theory\cite{Joan79},
self-consistent field theory\cite{Flee84}, perturbation
theory\cite{Walz94,Warr95}, direct simulations\cite{Dick94}, RG
theory\cite{Hank99,Schl01}, as well an interesting new ``overlap
approximation'' method\cite{Tuin01}. All these approaches show
significant deviations from ideal polymer behavior, but many questions
still remain. This is in contrast to binary HS colloid mixtures, where
the deviations from the AO potential can now be quantitatively
calculated with density functional theory
(DFT)\cite{Roth00,Gotz99,Roth01}, and the effects of non-ideality on
the phase behavior are fairly well understood\cite{Dijk98}.  The goal
of our paper is to derive a theory of similar accuracy for the
depletion potential induced by interacting polymers in a good solvent

Before we proceed, an important caveat is that the depletion potential
becomes less relevant for phase behavior at larger size ratios $q$,
since many-body interactions become increasingly
important\cite{Meij94,Loui99a,Dijk99}.  When $q>>1$, i.e.\ when the
colloids are much smaller than the polymers, other approaches, which
treat the polymers on a monomer level, are more relevant.  Examples
include integral equation techniques\cite{Fuch01}, scaling
theory\cite{deGe79a,deGe79}, or renormalization group theory
(RG)\cite{Eise00}.  In this paper we concentrate on the regime where
$R_g$ is of the order of $R_c$ or smaller.

We have recently performed a systematic study of the insertion
free-energy of a single colloidal particle in a bath of interacting
polymers\cite{Loui01e} (henceforth referred to as paper I), and found
important deviations from ideal polymer behavior.  In particular, we
found that the range of the  depletion layer decreases with increasing
density $\rho$, and that this effect is most pronounced in the dilute
regime, where $\rho \lesssim \rho*$; here $\rho* = \frac43 \pi R_g^3$
is the overlap density.  In the semi-dilute regime, where $\rho/\rho*
> 1$, we found that the effects of the interactions could be well
described by scaling theory\cite{deGe79}. Since these one-body effects
were not well captured by (non-additive) HS like models, we do not
expect straightforward extensions of the DFT techniques that work so
well for binary hard HS mixtures, to also perform well for two-body
depletion potentials in polymer-colloid mixtures.

To calculate the depletion potentials, we use a similar approach to
that used in paper I -- a combination of scaling theories and computer
simulations.  Since we found in paper I that the range of the
depletion layer decreases with increasing density, we expect a similar
trend for the range of the related depletion potential.  Throughout
this work we focus on the dilute and semi-dilute
regimes\cite{deGe79,Doi86} of the polymers, where the monomer density
$c$ is low enough for detailed monomer-monomer correlations to be
unimportant; the melt regime, where $c$ becomes appreciable, will not
be treated here.   Since our models are all athermal, we
set $\beta = 1/(k_B T) =1$.

The paper is organized as follows: The case of the depletion
interaction between two plates is discussed in detail in section II.
We show that it is closely related to the problem of determining the
surface tension $\gamma_w(\rho)$, which was solved in paper I.  Just
as was found for $\gamma_w(\rho)$, the depletion potential simplifies
in the semi-dilute regime. In section III, we discuss the depletion
interaction between two spheres.  We compare the results of direct
Monte Carlo (MC) simulations of the interaction between two HS
colloids induced by a bath of self avoiding walk (SAW) polymers to a
potential derived within the Derjaguin approximation\cite{Derj36}.
The Derjaguin approximation works much better than one would naively
expect because of a cancellation of errors related to the deformation
of polymers around a sphere. Using an extension of the Derjaguin
approximation, we derive a new semi-empirical depletion potential which
appears to be nearly quantitative for $q \leq 1$, the regime where
depletion potentials are most relevant for the phase behavior and
structure of polymer-colloid mixtures.  We derive the scaling behavior
with density, and find important deviations from the AO model and
other ideal-polymer theories.  We also calculate the depletion
potential between spheres within our new ``polymers as soft colloids''
coarse-graining scheme\cite{Loui00,Bolh01,Bolh01a,Bolh01b}, where each
polymer is represented by a single particle, interacting via a
density-dependent effective potential.  Here we again find good
agreement with the direct MC results for the dilute regime, but for
$\rho/\rho* > 2$, significant deviations are found.  In
section IV, we discuss the effect of polymer density on the virial
coefficients between two colloids, and relate this to phase behavior
of polymer-colloid mixtures.

\section{Depletion potential between two walls}

\subsection{Surface tension near a single wall and the
 depletion potential at contact}

Immersing a single hard wall (or plate) into a bath of non-adsorbing
polymers in a good solvent reduces the number of configurations
available for the polymers. This in turn results in an entropically
induced depletion layer $\rho(z)$ near the wall, discussed in more
detail, for example, in paper I. Associated with the creation of this
depletion layer is an interfacial free energy cost per unit area $A$,
i.e.\ a wall-fluid surface tension $\gamma_w(\rho)$, which typically
depends on the bulk density $\rho$, or equivalently the bulk chemical
potential $\mu$\cite{murho}.  Bringing two such walls, of area $A$,
together from an infinite distance apart to a distance $x$ where the
two depletion layers begin to overlap, changes the total free energy of
the polymer solution.  This change in the free energy or grand
potential $\Omega$ per unit area is called the depletion potential
$W(x)\equiv(\Omega(x) - \Omega(x=\infty))/A$\cite{murho}. When $x=0$,
i.e.\ when the two walls are brought into contact, the depletion
potential reduces to the simple form
\begin{equation}\label{eq2.1}
W(0) = - 2 \gamma_w(\rho),
\end{equation}
reflecting the fact that the two depletion layers are completely
destroyed.

In paper I, we used an extension of the Gibbs adsorption equation to
express the surface tension near a single wall as
\begin{equation}\label{eq2.2}
\gamma_w(\rho) = - \int_{0}^{\rho} \left( \frac{\partial
\Pi(\rho')}{\partial \rho'}\right) \hat{\Gamma}(\rho') d\rho'.
\end{equation}
The derivation of this equation can be found, for example,
in\cite{Mao97,Tuin01}.  Here $\Pi(\rho)$ is the osmotic pressure
of the polymer solution, and $\hat{\Gamma}(\rho)$ is the reduced
adsorption near a single wall, defined as:
\begin{equation}\label{eq2.3}
\hat{\Gamma}(\rho) = - \frac{1}{\rho}\frac{\partial
 (\Omega^{ex}/A)}{\partial \mu} = \int_{0}^\infty \left( \frac{\rho(z)}{\rho}
 -1\right) dz,
\end{equation}
where $\Omega^{ex}/A$ is the excess grand potential per unit area.  In
paper I, we used computer simulations of a SAW polymer solution in a
good solvent near a wall to calculate  $\rho(z)$ and
$\hat{\Gamma}(\rho)$ for several values of $\rho/\rho*$.  As discussed
in paper I, the SAW on a cubic lattice is a very good model for
polymers in a good solvent.  In the scaling limit, where the length
$L$ tends towards $\infty$, its properties are universal, and agree
with experiments on polymers in the same good solvent
regime\cite{deGe79,Doi86}.  For example, the radius of gyration scales
as $R_g \sim L^{\nu}$, where $\nu$ is the Flory exponent, taken to be
$\nu = 0.588$ in this paper.  By using an accurate fitting form which
takes into account the correct scaling behavior, we expressed
$\hat{\Gamma}(\rho)$ for all densities in the dilute and semi-dilute
regimes.  When this was combined with an RG expression for the
pressure, we were able to calculate the surface tension
$\gamma_w(\rho)$ as a function of density through
Eq.~(\ref{eq2.2}). Our results agreed very well with some recent RG
calculations\cite{Maas01}.

\begin{figure}
\begin{center}
\epsfig{figure=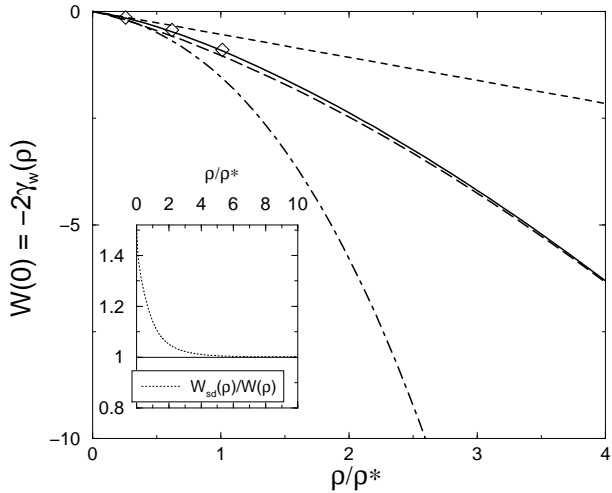,width=8cm}
\caption{\label{fig:W02-W0semi} Depletion potential between two walls
or plates at contact $W(0) = - 2 \gamma_w(\rho)$ per unit area
$R_g^2$.  The full lines result from Eq.~(\protect\ref{eq2.1}), while
the long-dashed lines come from the simpler semi-dilute scaling
expression (\protect\ref{eq2.7}).  The diamonds denote previously
published $L=100$ SAW simulation results\protect\cite{Bolh01}.  The
dashed line represents the simple ideal polymer form $W_{id}(0)$, given
by Eq.~(\protect\ref{eq2.10}), while the dot-dashed line shows the
naive improvement obtained by substituting the true polymer osmotic
pressure for the ideal polymer pressure.  These two Asakura-Oosawa
type approximations bracket the correct result.  The inset shows the
strength of the simpler semi-dilute scaling expression relative to the
the full potential at contact (dotted line).  The two coincide for
higher densities.  In the low density limit, the semi-dilute scaling
expression overestimates the true value at contact by a factor $1.5$.}
\end{center}
\end{figure}

 Here we use $\gamma_w(\rho)$ together with Eq.~(\ref{eq2.1}) to directly
calculate the depletion potential at contact, as shown in
Fig.~\ref{fig:W02-W0semi}. These results are also compared to direct
computer simulations of $L=100$ SAW polymers\cite{Loui00,Bolh01}.
Even though the $L=100$ simulations have not quite reached the scaling
limit, the well-depths at contact are remarkably well reproduced, a
result also found by Tuinier and Lekkerkerker\cite{Tuin01} with a
similar theory.

\subsection{A simple theory for the depletion potential}

The simplest approximation for the depletion potential at finite
separation $x$ would be a linear form with a slope equal to the
osmotic pressure.  This follows because $\Pi(\rho) x$ is the volume
term proportional to the work per unit area produced by the osmotic
pressure:
\begin{eqnarray}\label{eq2.6}
W(x)  = & W(0) + \Pi(\rho) x; & \,\,\,\, x \leq D_w(\rho)   \nonumber \\
W(x)  = & 0 \,\,\,\,\,\,\,\,\,\,\,\,\,\,\,\,\,\,\,\,\,\,\,\,\,\,\,\,\,\,\,\,\,\,\,\,\,\,\,\,\,\,\,\,\,\,\,
; & \,\,\,\, x > D_w(\rho)   
\end{eqnarray}
where the range is given by
\begin{equation}\label{eq2.6a}
D_w(\rho) = -\frac{W(0)}{\Pi(\rho)} = \frac{2 \gamma_2(\rho)}{\Pi(\rho)}.
\end{equation} 
  We note that this approximation is similar to that adopted by
Joanny, Leibler, and de Gennes who, in their pioneering
paper\cite{Joan79}, approximated the force between two plates as
constant for $x \leq \pi \xi(\rho)$ and zero for $x > \pi \xi(\rho)$,
where $\xi(\rho)$ is the correlation length, the relevant length-scale
in the semi-dilute regime\cite{deGe79,Doi86}. This also results in a
linear depletion potential.

\begin{figure}
\begin{center}
\epsfig{figure=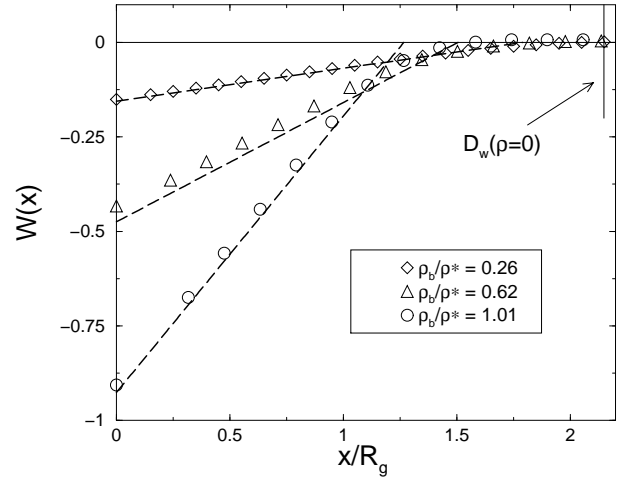,width=8cm}

\caption{\label{fig:vwall-simple} Depletion potential between two
walls or plates $W(x)$ per unit area $R_g^2$.  The symbols denote
simulations for $L=100$ SAW polymers, the straight lines are the
simple theory of Eq.~(\protect\ref{eq2.6}), which provides a near
quantitative fit to the simulation data.  The range
$D_w(\rho=0)\approx 2.15$
is shown as a vertical line.  Note how much the range decreases with
density, even for these results in the dilute regime.  }
\end{center}
\end{figure}

In Fig.~\ref{fig:vwall-simple} we compare our simple linear potential
to the direct SAW simulations.  The overall agreement is striking.
The only (small) deviation occurs at larger distances $x$ where the
true potential rounds off and develops a very small maximum before
going to zero\cite{zero}.  The range $D_w(\rho)$ of this simple
potential decreases with $\rho/\rho*$.  In fact, as shown in
Fig.~\ref{fig:range}, the largest relative rate of decrease occurs in
the dilute regime, so that at $\rho/\rho*=1$, the range is $D=1.25
R_g$, about $58\%$ of the low density limit of $D \approx 2.15 $.  In
general, the range of a potential is not always unambiguously
defined\cite{Noro00}.  Of course for our linear one it is, and this
simple definition would seem a very reasonable definition for the
range of the full depletion potentials depicted in
Fig.~\ref{fig:vwall-simple}.

\subsection{Scaling theory for the depletion potential in the semi-dilute regime}

\begin{figure}
\begin{center}
\epsfig{figure=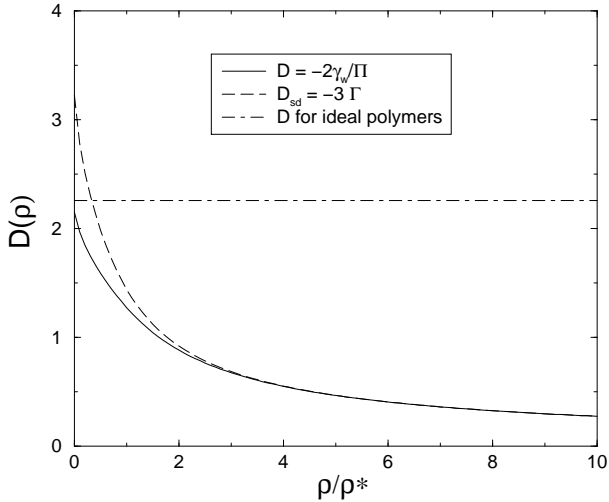,width=8cm}
\caption{\label{fig:range} Comparison of the range $D_w(\rho)$ of the
depletion potential with the simpler semi-dilute scaling expressions
$D_{sd}(\rho)$, both in units of $R_g$.  As density increases well
into the semi-dilute regime, the two expressions coincide.  In the low
density limit the semi-dilute expression is exactly $1.5$ times the
full expression.  Note how much the range differs from the density
independent result for non-interacting polymers $D_{id}=4/\sqrt{\pi}
\approx 2.26$ }
\end{center}
\end{figure}

Further simplifications occur in the semi-dilute regime. For example,
when the scaling forms\cite{deGe79} for the osmotic pressure, $\Pi
\sim \rho^{3 \nu/(3 \nu -1)}$ and the reduced adsorption $\hat{\Gamma}
\approx - \xi(\rho) \sim \rho^{-\nu/(3 \nu -1)}$ are used in
Eq.~(\ref{eq2.2}), then, as shown in paper I, the integral simplifies
and the potential at contact takes on the very simple form:
\begin{equation}\label{eq2.7}
W_{sd}(0) = 3 \Pi(\rho) \hat{\Gamma}(\rho) \sim \rho^{\frac{2 \nu}{3 \nu -1}} \approx \rho^{1.539}.
\end{equation}
  The linear depletion potential (\ref{eq2.6}) then reduces to:
\begin{eqnarray}\label{eq2.8}
W_{sd}(x) = & \Pi(\rho)\left(3 \hat{\Gamma}(\rho) +
x\right); & \,\,\,\, x \leq D_{sd} \nonumber \\ W_{sd}(x) = & 0
\,\,\,\,\,\,\,\,\,\,\,\,\,\,\,\,\,\,\,\,\,\,\,\,\,\,\,\,\,\,\,\,\,\,\,\,\,\,\,\,\,\,\,\,\,\,\,
; & \,\,\,\, h > D_{sd},
\end{eqnarray}
and  the range has simplified to
\begin{equation}\label{eq2.9}
D_{sd}(\rho) = - 3 \hat{\Gamma}(\rho) \sim \rho^{-\nu/(3 \nu -1)}
\approx \rho^{-0.770}.
\end{equation}
As can be seen in Figs~\ref{fig:W02-W0semi}~and~\ref{fig:range}, these
simple expressions for the well-depth and the range work remarkably
well in the semi-dilute regime, especially for $\rho/\rho* > 2$.

Deviations do occur for low densities since according to (\ref{eq2.1})
and (\ref{eq2.2}), $W(x=0) \approx 2 \Pi(\rho) \hat{\Gamma}(\rho)$ for
$\rho \rightarrow 0$, the same form as for ideal polymers.
Eq.~(\ref{eq2.7}) therefore overestimates the well depth by a factor
$1.5$, as can be seen in the inset of Fig.~\ref{fig:W02-W0semi}.  In
the same limit the range of the depletion potential reduces to the
functional form $D = -2 \hat{\Gamma}(\rho)$, so that Eq.~(\ref{eq2.9})
also overestimates the range at low densities by a factor $1.5$, as
can be seen in Fig.~\ref{fig:range}.

In the semi-dilute regime one can identify $\hat{\Gamma}(\rho) \approx
-\xi(\rho)$\cite{deGe79}.  Therefore the ansatz $D_w(\rho) = \pi
\xi(\rho)$, originally postulated by Joanny et al.\cite{Joan79}, is
very close to $D_{sd}(\rho) = - 3 \hat{\Gamma}$, the expression we
derived for the range of the  depletion potential  in the semi-dilute regime.
Their potential is therefore quite accurate in the semi-dilute regime,
but overestimates the range and the well-depth by a factor slightly
larger than $1.5$ in the dilute limit.

\subsection{Comparison with theories for non-interacting polymers between
two walls}


For completeness we compare the results obtained in the previous
section to those for ideal polymers, first obtained by Asakura and
Oosawa in 1954\cite{Asak54}.  Their (exact) depletion potential can be
quite accurately approximated by a simple linear form\cite{Bolh01}:
\begin{eqnarray}\label{eq2.10}
W_{id}(x) = & \rho\left(-\frac{4}{\sqrt{\pi}} + x\right); & \,\,\,\, x
\leq \frac{4}{\sqrt{\pi}} R_g \nonumber \\ W_{id}(x) = & 0
\,\,\,\,\,\,\,\,\,\,\,\,\,\,\,\,\,\,\,\,\,\,\,\,\,\,\,\,\,\,\,\,\,\,\,\,\,\,\,\,\,\,\,\,\,\,\,
; & \,\,\,\, x > \frac{4}{\sqrt{\pi}} R_g.
\end{eqnarray}
However, for interacting polymers, this is only true in the limit
$\rho \rightarrow 0$; the validity of this expression rapidly
deteriorates with increasing density. As shown in
Fig.~\ref{fig:W02-W0semi}, Eq.~(\ref{eq2.10}) underestimates the well
depth for all but the lowest densities, while, as shown in
Fig~\ref{fig:range}, it overestimates the range at all densities. One
might think that replacing the ideal pressure $\Pi(\rho) = \rho$ in
Eq.~(\ref{eq2.10}) by the pressure of an interacting polymer system
would bring an improvement for the well depth, even if this does not
improve the approximation for the range.  Instead, as shown in
Fig.~\ref{fig:W02-W0semi}, this naive approach leads to a severe
overestimate of the well-depth.  In fact, for the semi-dilute regime,
the contact value of this naive ``improvement'' would scale as
$W_{id}^{naive}(0) = (4/\sqrt{\pi})\Pi(\rho) \sim \rho^{3 \nu/(3
\nu-1)} \approx \rho^{2.309}$ instead of the correct $W(0) \sim
\rho^{2 \nu/(3 \nu -1)} \approx \rho^{1.539}$ scaling.  Even for
$\rho/\rho* = 1$ the differences are significant: $W(0) = -0.93$,
while $W_{id}(0) = -0.54$ and $W_{id}^{naive}(0) = -1.68$. For
$\rho/\rho* =10$, $W(0)=-25.4$, $W_{id}(0) = -5.4 $ and
$W_{id}^{naive}(0) = -208$!  In other words, the simple heuristic
ansatz is almost never a real improvement; for interacting
polymers, the Asakura-Oosawa potential between two plates is only
accurate at the very lowest of densities.

\section{Depletion interactions between spheres}

We now turn to  the polymer induced depletion
interaction between two hard spheres. Similarly to the case of two
walls, when the surfaces of two such spheres are brought to within a
distance where their depletion layers begin to overlap, the total free
energy of the system changes; this change again defines the depletion
potential.

\vspace*{-0.7cm}
\begin{figure}
\begin{center}
\epsfig{figure=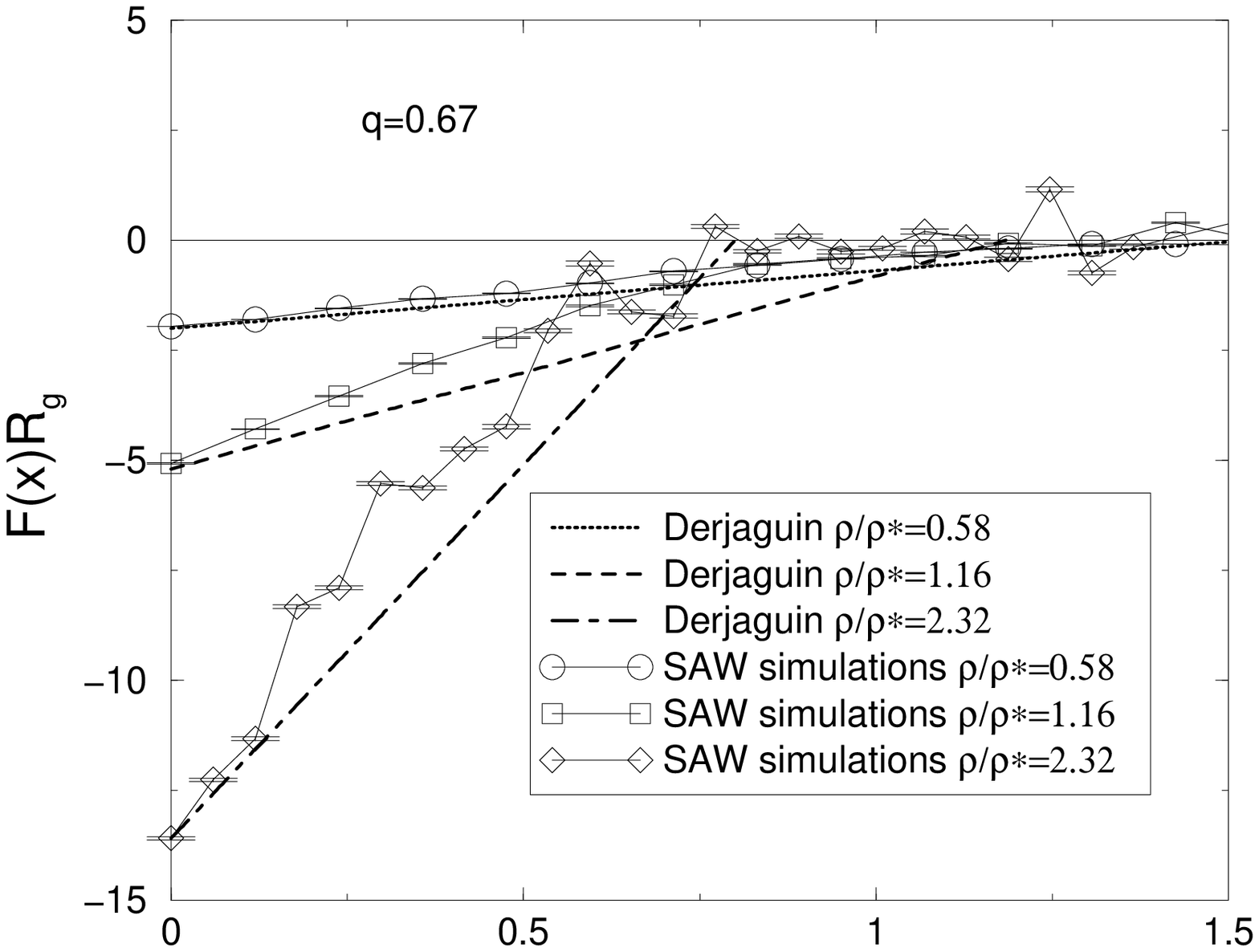,width=8cm}
\epsfig{figure=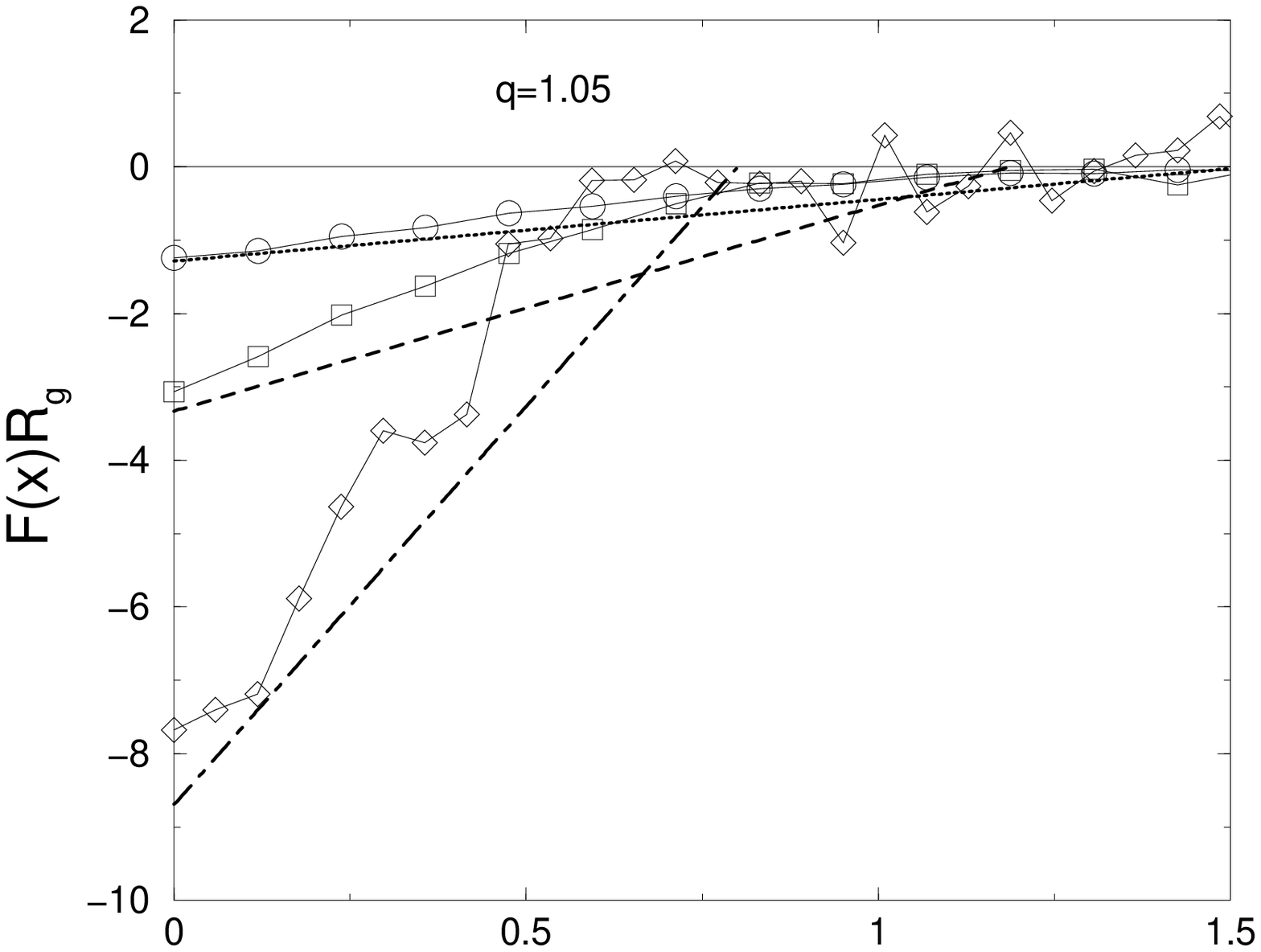,width=8cm}
\epsfig{figure=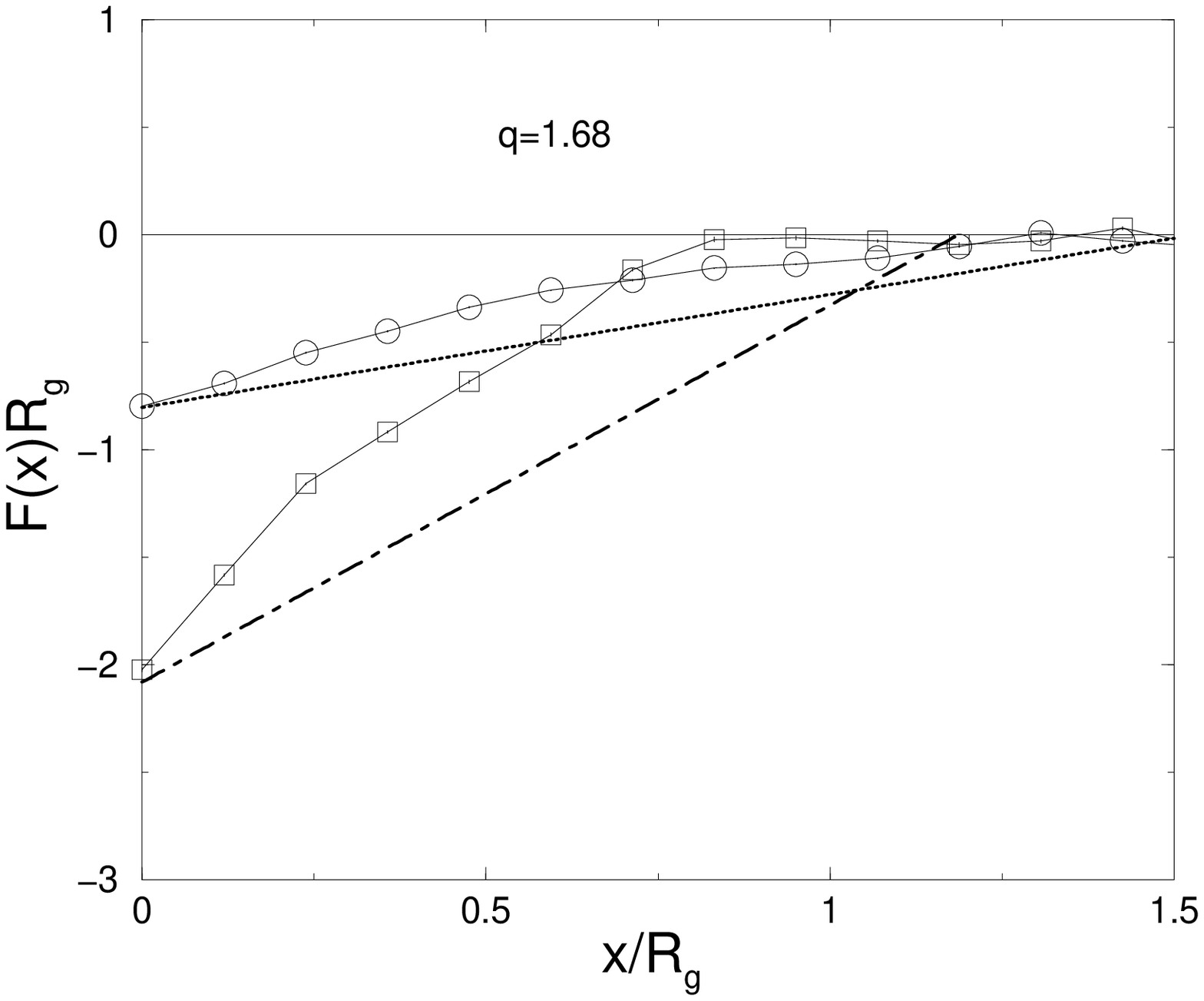,width=8cm}
\caption{\label{fig:Forces}Depletion forces between two spheres for
three size ratios $q$, as a function of $x$, the distance between the
surface of the two spheres.  The symbols denote $L=500$ SAW computer
simulations for three densities: $\rho/\rho*=0.58$ (circles),
$\rho/\rho*=1.16$ (squares), $\rho/\rho*=2.32$ (diamonds), the solid
lines are to guide the eye.  These are compared to results from the
Derjaguin approximation for the same size ratios and densities:
$\rho/\rho*=0.58$ (dotted lines), $\rho/\rho*=1.16$ (dashed lines),
and $\rho/\rho*=2.32$ (dot-dashed lines). }
\end{center}
\end{figure}

  Since we found in paper I that the surface tensions associated with
the polymer depletion layers around a single sphere show behavior
which differs from that of ideal polymers or that of the AO model, we
expect to find qualitative differences in the depletion potentials as
well. The trends are expected to be similar to those found for the depletion
potential between two walls, namely that the range should decrease and
the well depth should increase with increasing polymer concentration.

\subsection{Full SAW polymer simulations}\label{SAWdepletion}

To directly calculate the polymer induced depletion potential between
two spheres, we performed grand-canonical simulations of $L=500$ SAW
polymers on a lattice of size $240x150x150$, and computed the osmotic
pressure or force exerted on two hard spheres placed in the same
simulation box.  The configuration space of polymers was sampled in
the grand canonical ensemble with a combination of Configurational
Bias Monte Carlo and pivot moves\cite{frenkelbook}. The depletion
force on the spheres was obtained from the ratio of the acceptance of
virtual inward and outward moves of the spheres.

In Fig.~\ref{fig:Forces} these forces are shown for three different
size ratios $q=Rg/R_c$, namely $q=0.67$, $q=1.05$, and $q=1.68$.  For the
first two size ratios, we computed the forces at three densities,
$\rho/\rho*=0.58$, $\rho/\rho* = 1.16$, and
$\rho/\rho*=2.32$\cite{murho}, for $q=1.68$, this was only done for
the lower two densities.  As expected, for a fixed size ratio, the
range decreases, and the force at contact increases with increasing
polymer density.  For a given density, the range appears to contract
slightly with increasing $q$, as might be expected, since the polymers
can deform more readily around the smaller colloids (see the Appendix
for further discussion of this point).  In each simulation we keep the
size of the polymers fixed, so that the larger size ratios essentially
correspond to smaller spheres. The computational costs scales with the
size of the simulation box, which roughly sets the number of polymers
needed to achieve a density $\rho/\rho*$ in the accessible volume left by
the colloids.

\begin{figure}
\begin{center}
\epsfig{figure=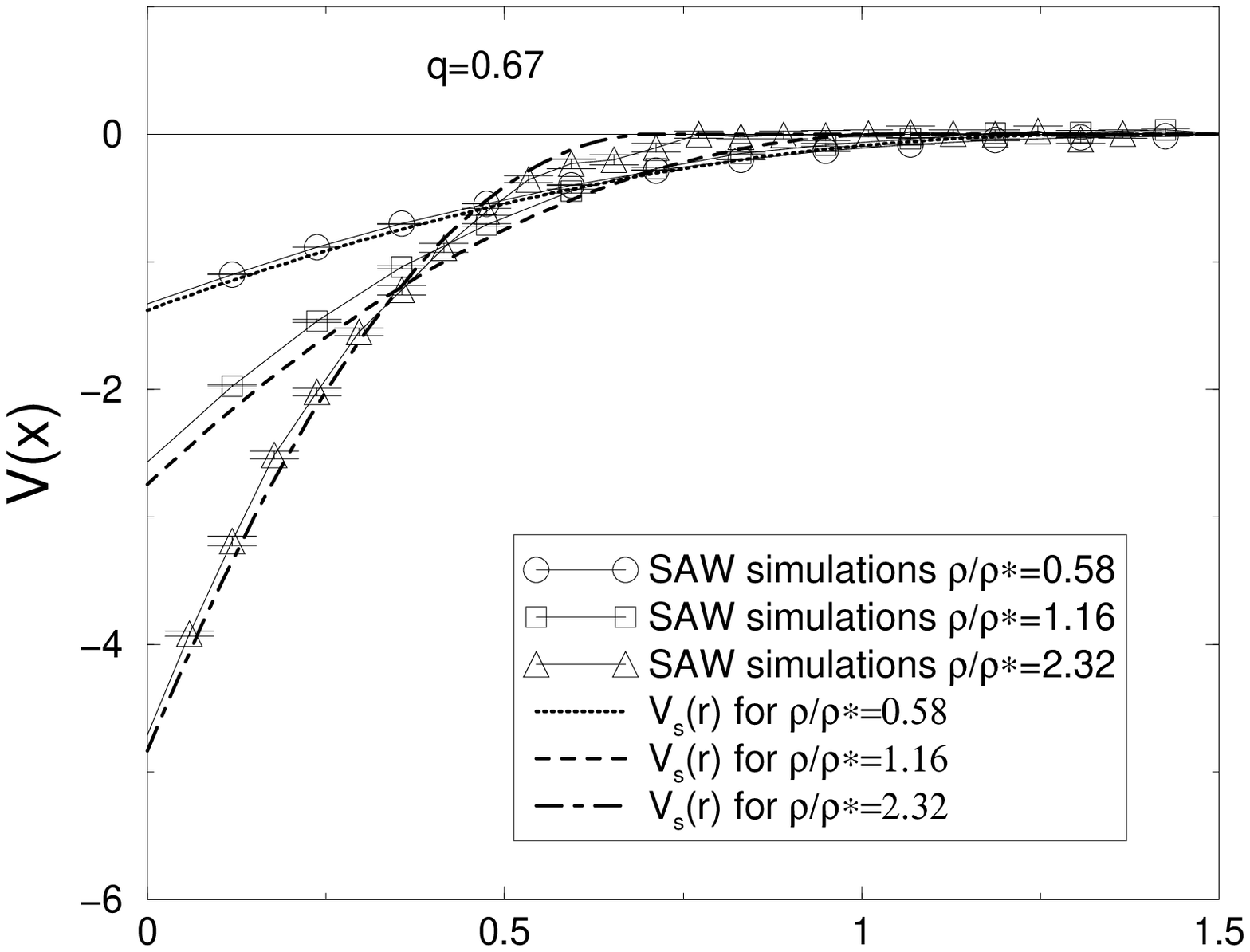,width=8cm}
\epsfig{figure=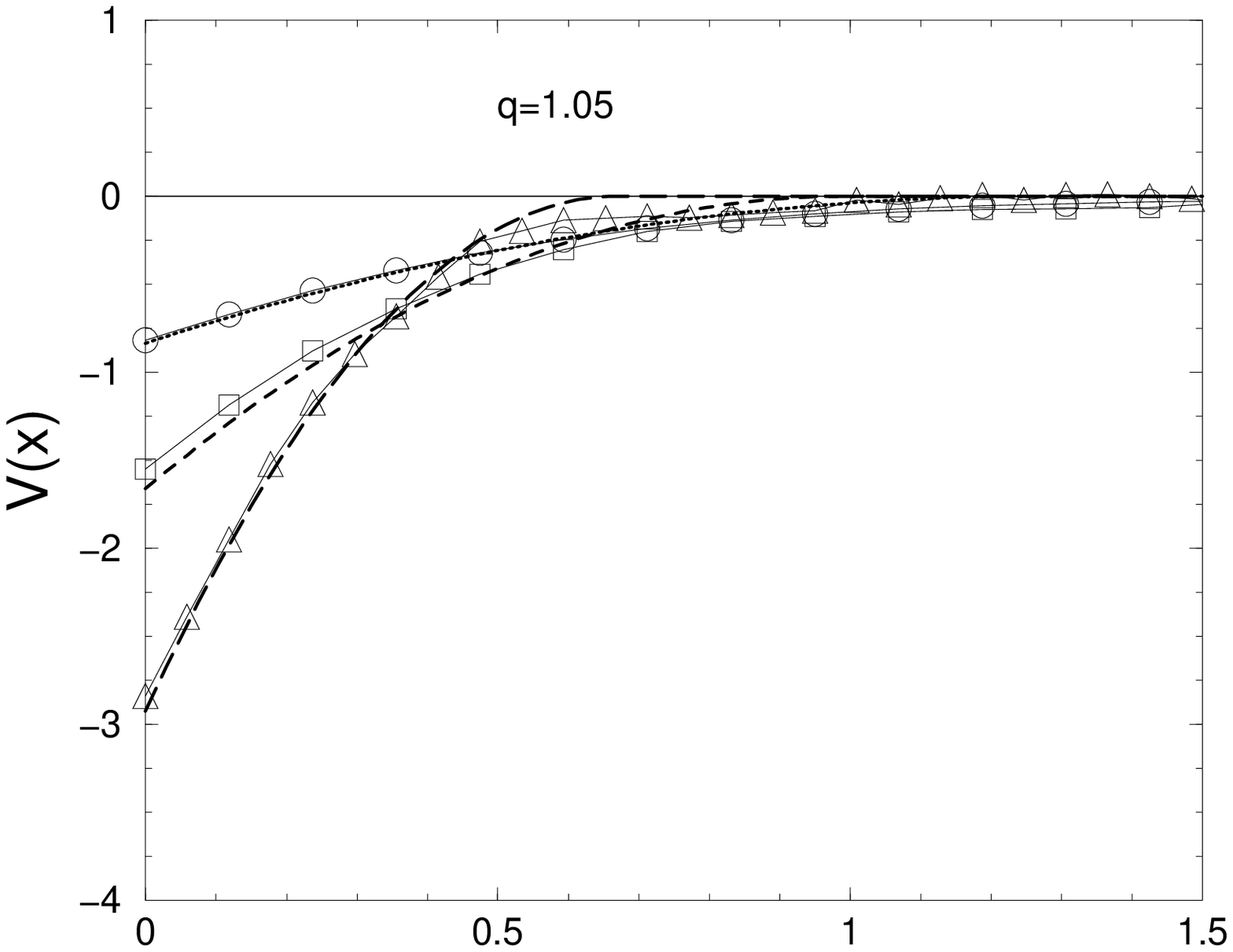,width=8cm}
\epsfig{figure=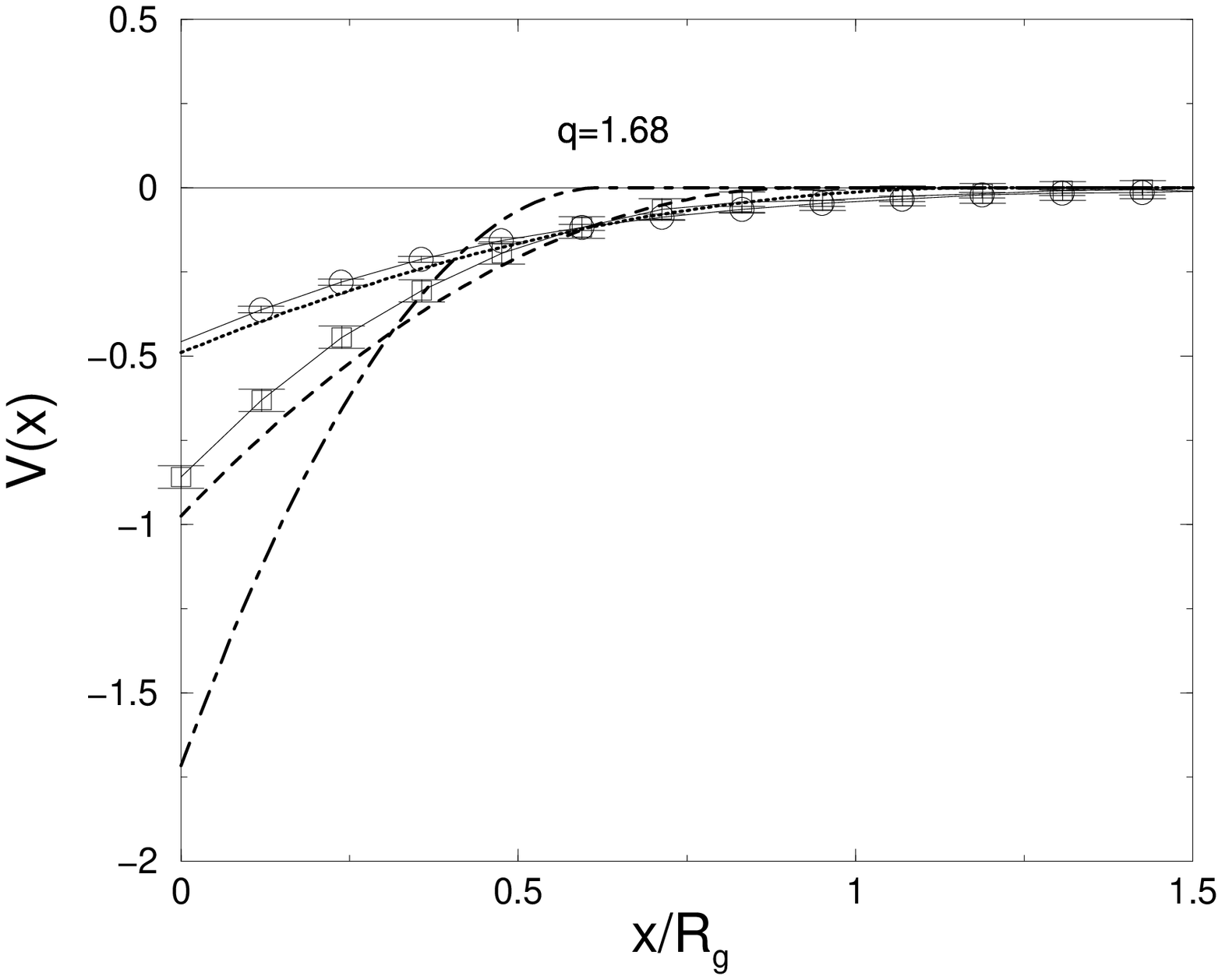,width=8cm}
\caption{\label{fig:vrad16-ejm-newpot} Depletion potentials between
two spheres for the same set of parameters as in
Fig.~\protect\ref{fig:Forces}.  These are compared to the
semi-empirical depletion potential $V_s(r)$ (Eq.~\protect\ref{eq3.21})
for the same size ratios and densities.}
\end{center}
\end{figure}

 The force can then be integrated to obtain the effective depletion
potential:
\begin{equation}\label{eq3.1}
V(x) = - \int_x^\infty F(y) dy.
\end{equation}
Results are shown in Fig.~\ref{fig:vrad16-ejm-newpot} for the same
parameters as in Fig.~\ref{fig:Forces}.  Because the curves are
integrated, they appear smoother than the forces. There may still be
some residual error in the potentials, which may explain why the range
for the highest density seems slightly larger for the potential than
for the force. Also, the errors are likely too large to decide whether
or not there is a repulsive bump in the potential. If it does exist,
it is probably very small\cite{Guch00}.  This is in contrast to
hard-core systems, where such bumps can be
pronounced\cite{Walz94,Mao95,Loui01f}.  The weakness or absence of a
repulsive barrier in the case of interacting polymers can probably be
traced to the very low monomer concentration $c$.  For shorter polymers,
whose behavior deviates significantly from the $L \rightarrow \infty$
scaling limit, and where a substantial monomer density $c$ can more
easily be achieved, such repulsive bumps might occur more
readily\cite{Brou00}.

As expected, the range decreases while the force at contact and the
potential well-depth increase with increasing polymer density.  For a
fixed density and polymer size $R_g$, the force at contact and the
well-depth decrease as the spheres become smaller (or $q$ becomes
larger). The same effect occurs for ideal polymers\cite{Meij94},
where it has a simple geometric origin: the volume of a depletion layer of
a given width $\sim R_g$ decreases with decreasing size of the hard spheres.

\subsection{The Derjaguin approximation for interacting polymers}

In the case of two walls, the well-depth at contact has a clear
interpretation in terms of the complete destruction of two depletion
layers, and can therefore be expressed as a simple function of the
wall-polymer surface tension $\gamma_w(\rho)$, as shown in
Eq.~(\ref{eq2.1}).  For two spheres, the depletion layers do not
completely overlap, and the depletion potential at contact is
related to the surface tension of  polymers surrounding two spheres
in a dumbbell type configuration.  A simple expression for the well
depth at contact in terms of the surface tension $\gamma_s(\rho)$ 
is therefore less obvious.

One way to make contact with the two wall case is to use the Derjaguin
approximation\cite{Derj36}, which relates the force between two spheres of
radius $R_c$ to the potential between two walls, $W(x)$, in the
following way:
\begin{equation}\label{eq3.2}
F^{Derj}(x) = \pi R_c W(x),
\end{equation}
where $x$ is the distance between the surfaces of the two spheres.  In
principle this approximation should only be accurate for very small
size ratios $q$, i.e.\ for very large colloids.  However, as shown in
Appendix A, we expect there to be a cancellation of errors, related to
the deformation of the polymers around spherical particles, which
makes the Derjaguin approximation work much better than one would
naively expect.  This is confirmed in Fig.~\ref{fig:Forces}, where the
Derjaguin expression for the force, taken from Eq.~(\ref{eq2.6}) and
Eq.~(\ref{eq3.2}), is shown to be a surprisingly good approximation.
It is most accurate for the smallest $q$, as expected, but even for
$q=1.68$, where normally one would not expect the Derjaguin
approximation to be useful at all, it is still reasonable.

\begin{figure}
\begin{center}
\epsfig{figure=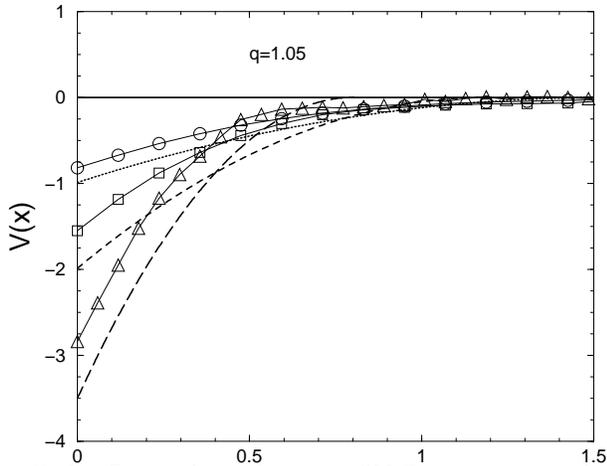,width=8cm}
\caption{\label{fig:vrad16-ejm-derj} Comparison of the $L=500$ SAW
computer simulations for the depletion potential with the Derjaguin
approximation of Eq.~(\protect\ref{eq3.10}) for $q=1.05$.
The symbols are the same as in Fig.~\protect\ref{fig:Forces}.
}
\end{center}
\end{figure}

By combining Eq.~(\ref{eq2.6}) with Eqs.~(\ref{eq3.1}) and
(\ref{eq3.2}) we obtain the following Derjaguin expression for the
depletion potential between two spheres
\begin{equation}\label{eq3.10}
V^{Derj}(x) = -\frac{\pi}{2} R_c \Pi(\rho) \left(D_w(\rho) - x
\right)^2
\end{equation}
for $0 \leq x \leq D_w(\rho)$; $V^{Derj}(x) =0$ for $x>
D_w(\rho)$. By the nature of the Derjaguin approximation, $D_w(\rho)$
is the same range as found for two walls in Eq.~(\ref{eq2.6a}).  This
simple potential is compared in Figs.~\ref{fig:vrad16-ejm-derj} to the
direct $L=500$ SAW simulation results for $q=1.05$.  The
correspondence is surprisingly good given the large size ratio $q$,
although it is not quantitative as in the wall-wall case. Similar
results are found for the two other size ratios. Overall, the
Derjaguin approximation overestimates the well-depth, a consequence of
the slightly longer ranged  forces found in
Fig.~\ref{fig:Forces}. Again, the cancellation of errors found for the
simpler AO model in Appendix A helps explain why Eq.~(\ref{eq3.10})
works reasonably well in a regime where the Derjaguin approximation
would normally break down.

\subsection{The Derjaguin approximation in the semi-dilute regime}

We now turn to the scaling behavior of the depletion potential in the
semi-dilute regime.  By using the simple expression for the range
$D_{sd}(\rho)$ given in Eq.~(\ref{eq2.9}), the Derjaguin approximation
for the depletion potential~(\ref{eq3.10}) reduces to
\begin{equation}\label{eq3.11}
V_{sd}^{Derj}(x) = - 
\frac{\pi}{2} R_c \Pi(\rho) \left[
3\hat{\Gamma}(\rho) + x \right]^2
\end{equation}
for $x \leq D_{sd}(\rho)$; $V_{sd}^{Derj}(x) =0$ for $x >
D_{sd}(\rho)$.  For two walls, the simplified expression (\ref{eq2.9})
for the range overestimates the true range by a factor $1.5$ in the
low-density limit.  Here the overestimate is slightly larger, since,
as shown in the appendix, the deformation of the polymers around a
single colloid reduces the depletion layer width for decreasing sphere
size $R_c$.  Similarly the well-depth at contact is also overestimated
in the low density limit, but now by a factor $\lim_{\rho\rightarrow
0} V_{sd}^{Derj}(0)/V^{Derj}(0) = 2.25$.  In the semi-dilute regime
the two expressions come closer for increasing density.  For example,
at $\rho/\rho*=2$ the overestimate at contact is only
$V_{sd}^{Derj}(0)/V^{Derj}(0)= 1.09$, while for $\rho/\rho*=4$ the two
expressions are within $1 \%$ of each other.  If in addition we assume
that $\hat{\Gamma}(\rho) \approx -\xi(\rho)$ then the expression in
Eq.~(\ref{eq3.11}) is again very similar to the form proposed by
Joanny {\em et al.}\cite{Joan79} for two spheres ($3$ being replaced
by $\pi$).  Our arguments therefore provide an a-posteori
justification of the validity of their potential for the semi-dilute
regime.  In the dilute regime it will overestimate the attraction in
the same way as Eq.~(\ref{eq3.11}) does.

\begin{figure}
\begin{center}
\epsfig{figure=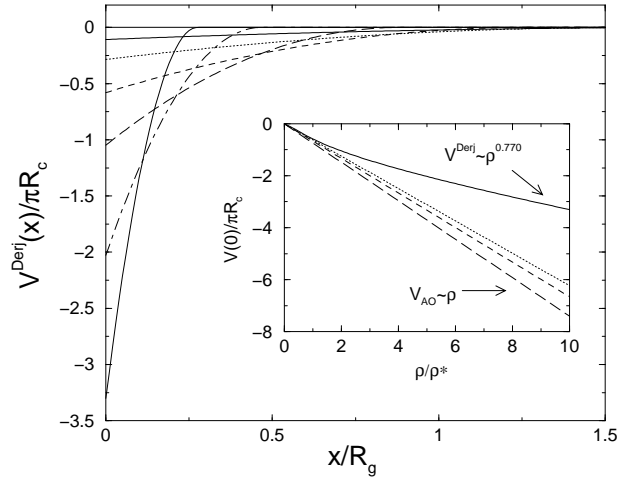,width=8cm}
\caption{\label{fig:Vderj} Scaled depletion potential
$V^{Derj}(x)/(\pi R_c)$ for interacting polymers.  From top to bottom
the densities are $\rho/\rho* = 0.2$, $0.5$, $1$, $2$, $5$ and $10$
respectively.  In the semi-dilute regime the well-depth increases like
$\rho^{0.770}$ while the range decreases as $\rho^{-0.770}$. Inset:
Scaled well depth at contact $V^{Derj}(0)/(\pi R_c)$ for the
interacting polymers (solid line), compared to the scaled well depth
$V_{AO}(0)/(\pi R_c)$ for the AO potential of
Eq.~(\protect\ref{eqA.1}) with $R_{AO} = R_{AO}^{eff}$ given by
Eq.~(\protect\ref{eqA.4}). The scaled AO potentials are shown for $R_c
= 10 R_g$, (dotted line), $R_c = 2 R_g$ (dashed line) and $R_c = 0.5
R_g$ (long-dashed line).  In contrast to the Derjaguin approximation
expressions, the AO potential does not satisfy prefect scaling with $R_c$.
The well-depth scales linearly in $\rho$.  The range is $2
R_{AO}^{eff}$, and is independent of $\rho$.  }
\end{center}
\end{figure}

Given that we now have a reasonably accurate depletion potential
between two spheres, namely Eq.~(\ref{eq3.10}), it seems fruitful to
examine how this expression varies with density.  We do this in
Fig.~\ref{fig:Vderj} for a number of densities in the dilute and the
semi-dilute regimes.  Since within  the Derjaguin approximation the
depletion potential (\ref{eq3.10}) always scales with $R_c$, the
curves in the plot can be used for all size ratios. Of course one
should keep in mind that the Derjaguin approximation becomes
progressively less accurate for decreasing $R_c/R_g$. With this caveat in
mind, the well-depth at contact for polymers in a good solvent scales
as:
\begin{equation}\label{eq3.12}
V_{sd}^{Derj}(0) \sim \rho^{\nu/(3 \nu -1)} \approx
\rho^{0.770}
\end{equation} in
the semi-dilute regime. This is exactly half the exponent with which
the well-depth at contact scales for two walls in the semi-dilute
regime (see Eq.~(\ref{eq2.7})), which can be understood as follows:
Within the Derjaguin approximation the force scales as $F(0) =\pi R_c
W(0) \sim \rho^{2\nu/(3 \nu -1)}$; the potential is then obtained
through Eq.~(\ref{eq3.1}) by integrating this force over a range
$D(\rho) \sim \rho^{-\nu/(3 \nu -1)}$. These two effects result in the
scaling seen in Eq.~(\ref{eq3.12}). On the other hand, by nature of
the Derjaguin approximation, the range is the same as that shown in
Fig.~\ref{fig:range} for the wall-wall case, i.e. it decreases as
$D_w(\rho) \sim \rho^{-0.770}$.

An expression  similar to Eq.~(\ref{eq3.11}) has been derived by Fleer,
Scheutjens and Vincent (FSV)\cite{Flee84}.  They replace $R_{AO}$ in
the full AO potential (\ref{eqA.1}) by $2 \hat{\Gamma}(\rho)$, and
the density $\rho$ by $\Pi(\rho)$.  In its original
applications\cite{Vinc88}, the FSV theory was used with a self
consistent field theory to calculate $\hat{\Gamma}(\rho)$, and a
simple Flory Huggins prescription for $\Pi(\rho)$.  This leads to
$\hat{\Gamma}(\rho) \sim \rho^{-\frac12}$ scaling behavior which is
more appropriate for polymers near the theta point\cite{deGe79}.
However, one could easily generalize the FSV approach and include the
$\hat{\Gamma}(\rho)$ and $\Pi(\rho)$ appropriate for polymers in a
good solvent.  For the semi-dilute regime the FSV approach would then
lead to an underestimate of the range by a factor $2/3$, while for
small $q$, where the Derjaguin approximation is quite accurate, it
would underestimate the attraction at contact by a factor of about
$4/9$.  On the other hand, in the dilute regime, this approach should
perform better, although for increasing $q$ it is expected to
overestimate the attraction at contact in a very similar way to the
effects seen for the AO model in Fig.~\ref{fig:VAO-Derj} (a).  This is
because the FSV model ignores the deformation of polymers near a
spherical surface.  It would be better to use the $D(\rho)$
appropriate for a spherical particle; this results in a potential
similar in spirit to an interesting recent proposal by Tuinier and
Lekkerkerker\cite{Tuin01b}.

\subsection{An accurate semi-empirical depletion potential}

\begin{figure}
\begin{center}
\epsfig{figure=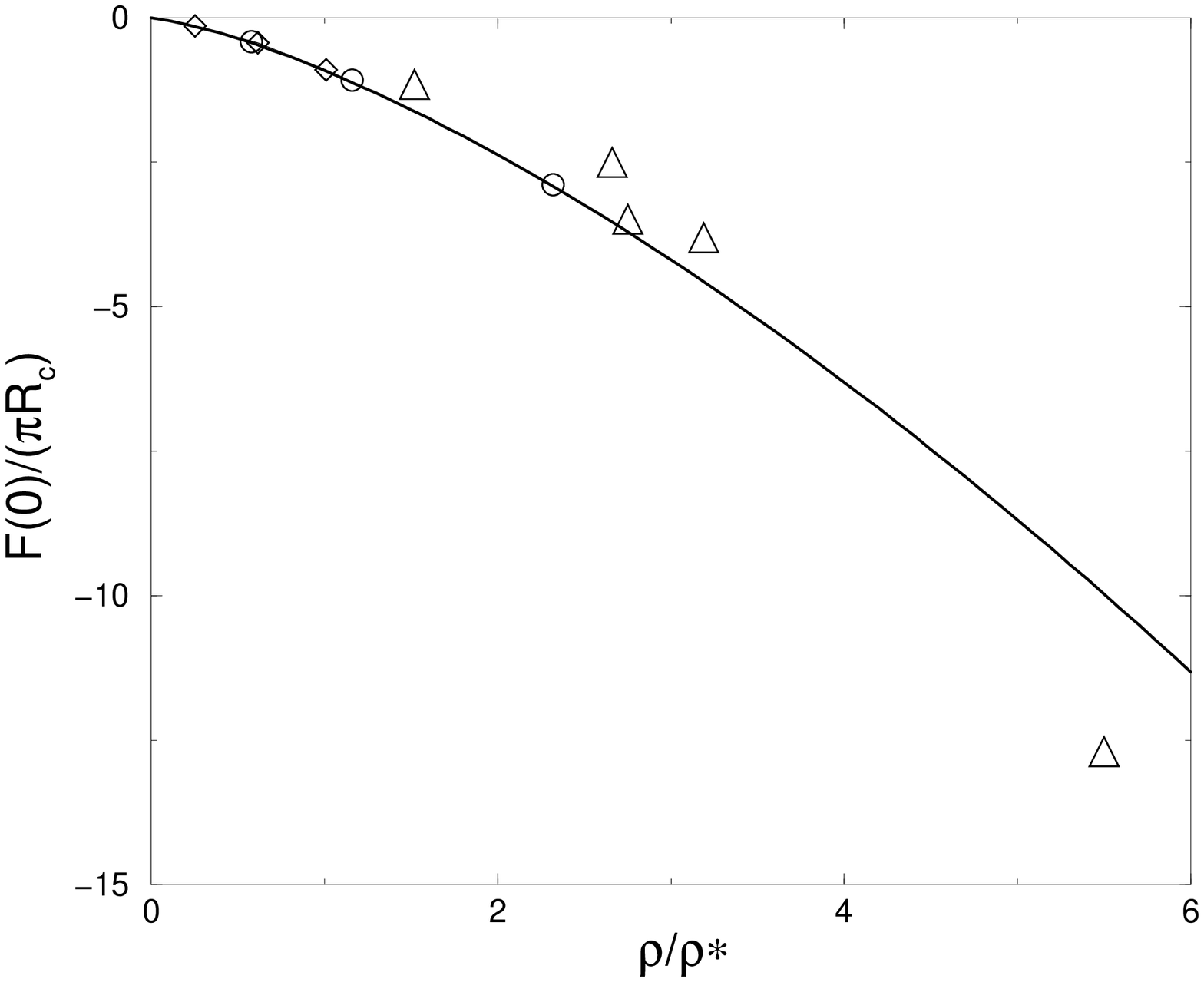,width=8cm}
\end{center}
\caption{\label{fig:F0}} Comparison of the force at contact with the
Derjaguin expression from Eq.~(\protect\ref{eq3.2}).  The circles are
taken from the SAW simulations depicted in
Fig.~\protect\ref{fig:Forces}.  On the scale of this graph, the three
size ratios lead to virtually identical results at each density.  The
triangles denote $F(0)/(\pi R_c)$ from the fluctuation bond model
simulations of Yethiraj and Dickman\protect\cite{Dick94}.  The
diamonds are from previous $L=100$ SAW simulations of $W(0)$, the
interaction between two plates\cite{Loui00,Bolh01}. The curve denotes
$W(0) = - 2 \gamma_w(\rho)$, the expected result for the Derjaguin
approximation.
\end{figure}

As can be seen in Fig.~\ref{fig:Forces}, the Derjaguin approximation
seems to be particularly accurate for the force at contact. In fact,
when we plot $F(0)/(\pi R_c)$ v.s.\ $\rho/\rho*$ in Fig.~\ref{fig:F0},
the results for the three size ratios are very close to each other at
each of the densities studied, and their value is well described
by $W(0) = - 2 \gamma_w(\rho)$, as expected for the Derjaguin
approximation (see Eq.~\ref{eq3.2}).  Furthermore, we analyzed the
computer simulation data of Dickman and
Yethiraj\cite{Dick94,Dickcompare}, which used the fluctuation bond
model (FBM) for size ratios ranging from $q=1.3$ to $q=3.5$, and find
similar agreement in Fig.~\ref{fig:F0}.  Keep in mind that the FBM
results are for shorter chains ($L=20 - 100$), and that the error
bars appear to be larger than those in our simulations.  Since for
$q=3.5$ the Derjaguin approximation (\ref{eq3.2}) should not be valid
at all, this does suggest that the expression
\begin{equation}\label{eq3.3}
F(0) = \pi R_c W(0) = - 2 \pi R_c \gamma_w(\rho),
\end{equation}
may hold because of an (approximate) sum-rule, and that the agreement
is not fortuitous.

The surprising accuracy of expression~(\ref{eq3.3}) for the force at
contact can be exploited to derive a more accurate effective potential
than that found by the direct application of the Derjaguin
approximation.  We assume the same linear force between two spheres as
found in the previous two sections, but now with a modified range
\begin{equation}\label{eq3.20}
D_s(\rho) = D_w(\rho) \frac{R_{AO}^{eff}(R_c)}{R_g},
\end{equation}
where the effective shrinking of the range due to the deformation of
the polymers around the HS has been taken into account via
Eq.~(\ref{eqA.4}).  Although this deformation effect is
calculated for ideal polymers, we expect it to be a good first
approximation for the {\em relative} deformation of interacting polymers
around a sphere.  The resulting simple linear expression for the force
can now be integrated up via Eq.~(\ref{eq3.1}) to give the depletion
potential:
\begin{equation}\label{eq3.21}
V_s(x) = \frac{\pi}{2} R_c W(\rho) D_s(\rho)\left(1 -
\frac{x}{D_s(\rho)}\right)^2
\end{equation}
for $x \leq D_s(\rho)$. $V_s(x) = 0$ for $x > D_s(\rho)$.  Our new
semi-empirical potential works remarkably well, as can be seen in
Fig.~\ref{fig:vrad16-ejm-newpot} for the three size ratios.  Even for
$q=1.68$, corresponding to the smallest colloids, this potential is
significantly better than the Derjaguin expression~(\ref{eq3.10}),
suggesting that Eq.~(\ref{eq3.21}) can be considered a nearly
quantitative representation of the depletion potential induced between
two spheres by SAW polymers for a surprisingly wide range of $q's$.
Since the semi-empirical potential~(\ref{eq3.21}) reduces to the
regular Derjaguin form~(\ref{eq3.10}), if $D_s(\rho)$ is replaced by
$D_w(\rho)$, its scaling properties with density should be the same as
those discussed in the previous subsections.

\subsection{Comparison with theories for non-interacting polymers}

It is interesting to compare the results for interacting polymers with
the results for non-interacting polymers.  Firstly, just as was found
for two walls, the range for the depletion potential is independent of
density for non-interacting polymers.  This is true both for the
simpler AO model\cite{Asak58}, as well as for more sophisticated
theories and computer simulations\cite{Meij94,Tria99,Tuin00}.
For a given $R_g$, using a theory based on non-interacting
polymers will always overestimate the range compared to the
interacting case.  Since for our accurate semi-empirical potential the
range $D_s(\rho)$ is related to $D_w(\rho)$ by a density independent
factor related to $q$, the {\em relative} overestimate in the range
for the spherical case should be close to that found when comparing a
non-interacting theory to an interacting one for the range of the
depletion potential between two plates, as done in
Fig.~\ref{fig:range}.

For the case of two walls the absolute value of the well-depth at
contact given by ideal polymers or the AO model was {\em
underestimated} w.r.t.\ the interacting case.  Within the Derjaguin
picture of Eq.~(\ref{eq3.2}), this implies that non-interacting
polymer models will underestimate the force at contact $F(0)$ for two
spheres in a bath of interacting polymers by a similar factor to that
shown for $W(0)$ in Fig.~\ref{fig:W02-W0semi}.

In the inset of Fig.~\ref{fig:Vderj} we compare the scaled well-depth
at contact $V^{Derj}(0)/(\pi R_c)$ for the Derjaguin approximation
with that of the AO model, given in Eq.~(\ref{eqA.1}), where an
effective range parameter $R_{AO}^{eff}$, given by Eq.~(\ref{eqA.4})
in Appendix A, was used to take into account the deformation of the
polymers around the colloids.  For a given $R_c$, the well-depth
$V_{AO}(0)$ scales linearly with $\rho$, which is now an {\em
overestimate} compared to the result for interacting polymers.  The
origin of this behavior can be easily understood: while the strength
of the force is strongly underestimated by the AO model, the range is
overestimated.  Therefore the integral in Eq.~(\ref{eq3.1}) shows a
compensation of errors.  This explains why, at low densities, the AO
approximation well depth is very close to the interacting polymer
result. For example for $q=0.1$ we find for $\rho/\rho*=1$ that
$V_s(0)/V_{AO}(0) \approx 0.93$ while for $\rho/\rho*=10$ this ratio
is still only $0.56$.  Attempting a naive improvement by replacing
$\Pi(\rho)=\rho$ in the AO expression by the true pressure of a
polymer solution results in a worse approximation for all densities.

\subsection{Polymers as soft colloids}

When modeling polymer-colloid mixtures, the colloids are usually
treated as single  particles.  It thus seems natural to
attempt the same for the polymers.  We have recently succeeded in
modeling linear polymers as ``soft
colloids''\cite{Loui00,Bolh01,Bolh01a,Bolh01b,Loui02a}, where each
polymer is replaced by a single particle centered on its CM.
Different coils interact via an effective pair potential acting
between their CM.  These effective interactions $v(r;\rho)$ between
the CM of the polymers were extracted from the radial distribution
functions $g(r)$ with an Ornstein Zernike inversion technique.  This
procedure is justified by a theorem which states that for any given
$g(r)$ and $\rho$ there exists a single unique pair-potential which
reproduces that radial distribution function\cite{Hend74,Reat86}.  Our
$v(r;\rho)$ have a near Gaussian shape, with an amplitude of order $2
k_B T$ and a range of the order of the radius of gyration $R_g$.  For
$\rho/\rho* \leq 2$ accurate analytic forms as a function of $r/R_g$
and $\rho/\rho*$ are available\cite{Bolh01b}.

 Our input $g(r)$'s were generated by computer simulations of $L=500$
SAW polymers on a cubic lattice, and so the density dependent
effective potentials we derived will by definition reproduce the same
CM pair structure as found in the underlying SAW polymer system.  When
used in conjunction with the compressibility equation\cite{Hans86},
these potentials also provide a very good representation of the equation
of state\cite{Bolh01,Bolh01b}.

A similar coarse-graining procedure was followed to describe polymers
near a flat wall\cite{Loui00,Bolh01}, where even ideal polymers form a
depletion layer. If these were to be modeled as single particles,
there would be no polymer-polymer interaction, but there would still
be an effective polymer-wall interaction $\phi(z)$.  We used direct simulations
of $L=500$ SAW polymers to extract the density profile $\rho(z)$ near
a hard wall.  The $\phi(z)$'s which reproduce the density profile
at each given bulk density $\rho$ were calculated using a
wall-Ornstein-Zernike procedure\cite{Loui00,Bolh01}.  Again accurate
analytic fits for densities $\rho/\rho^* \leq 2$ are
available\cite{Bolh01b}.

Since our effective polymer-polymer potentials provide a very accurate
representation of the pressure $\Pi(\rho)$, while the polymer-wall
interactions are constrained to reproduce the correct density profile
$\rho(z)$, and therefore the correct adsorption $\hat{\Gamma}(\rho)$,
Eq.~(\ref{eq2.2}) implies that our soft-colloid approach will reproduce the
correct wall-fluid surface tension $\gamma_w(\rho)$.  This explains why our
soft-colloid approach reproduces the correct value for
$W(0$)\cite{Loui00,Bolh01}, the depletion potential at contact
between two walls, since Eq.~(\ref{eq2.1}) implies that this can be
expressed completely in terms of $\gamma_w(\rho)$.  Similarly the
slope of the potential is given by the osmotic pressure
$\Pi(\rho)$ which is also accurately represented in the
soft-colloid approach.  Deviations w.r.t.\ direct simulations were
only observed for larger distances $z$, where the soft-colloid
approach produced a more pronounced repulsive bump in the depletion
potentials.\cite{Loui00,Bolh01}.

\begin{figure}
\begin{center}
\epsfig{figure=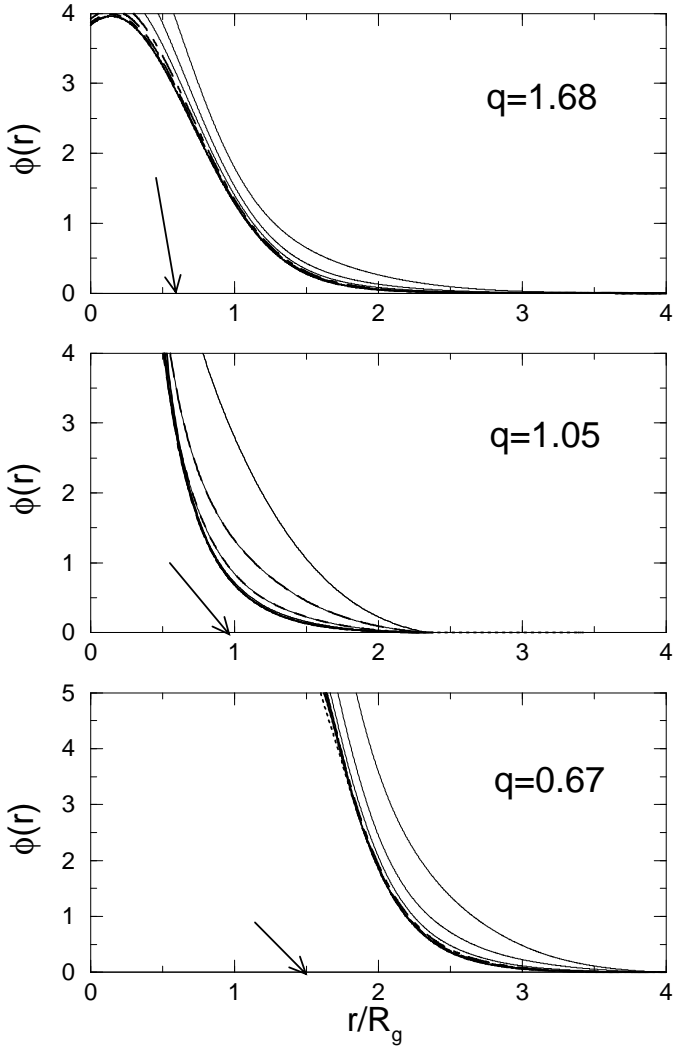,width=9cm}
\caption{\label{fig:vrsphereall} The effective sphere-polymer
potential $\phi(r)$ as a function of $r$, the distance to the centre
of the sphere for increasing polymer concentration (from left to
right). These $\phi(r)$ are derived via an Ornstein-Zernike inversion
technique such that they reproduce the CM density profiles around each
sphere\protect\cite{Bolh01b}. The arrows denote the hard core radius
of the spheres.  }
\end{center}
\end{figure}

To calculate the depletion potential between spheres, we need to first
derive the effective sphere-polymer CM potential $\phi(r)$ which would
exactly reproduce the density profile $\rho(r)$ around a single sphere
of radius $R_c$.  This can again be done using an Ornstein-Zernike
(OZ) technique to invert the density profiles\cite{Bolh01b}. These
density profiles were calculated in paper I, where we found that the
range of the profiles shrinks with increasing density. The CM profiles
can penetrate into the spheres because the polymers can deform around
them, an effect which becomes more pronounced with decreasing sphere
size (this doesn't happen in the monomer representation of course).
Similarly, the effective potentials $\phi(r)$, derived with our
inverse OZ approach, become relatively softer with decreasing $R_c$,
as shown in Fig.~\ref{fig:vrsphereall}\cite{Bolh01b}.  For a fixed $q$
the potentials become more repulsive for increasing density, just as
was found for the polymer-wall case\cite{Bolh01}.  Note that this
behavior is the opposite to that of the density profiles.  Again,
because these potentials are constrained to give the correct density
profiles and related adsorptions, this soft colloid picture should also
correctly reproduce the surface tension $\gamma_s(\rho)$ for a single
sphere immersed in a bath of interacting polymers.  The same should
hold for the related one-body insertion free energies
$F_1^{int}(\rho)$ described in paper I.

\subsubsection{Direct simulations of depletion potentials for polymers
as soft colloids}

\begin{figure}
\begin{center}
\epsfig{figure=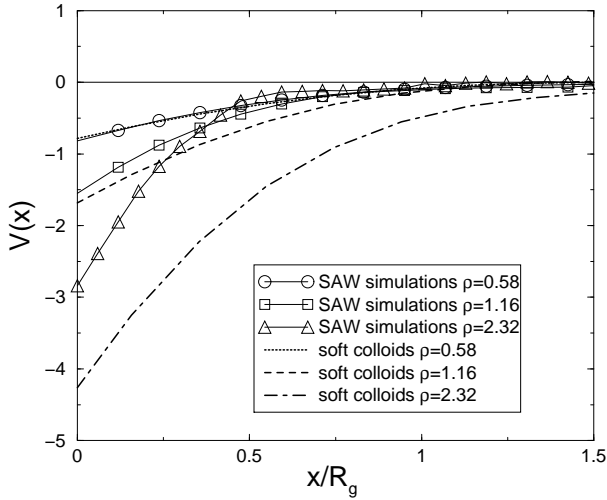,width=8cm}
\caption{\label{fig:vrad16-ejm-eff} Comparison of the direct SAW
simulations of the depletion potential for $q=1.05$ to direct
simulations within the polymers as soft colloids approach.  }
\end{center}
\end{figure}

\begin{figure}
\begin{center}
\begin{minipage}{8cm}
\epsfig{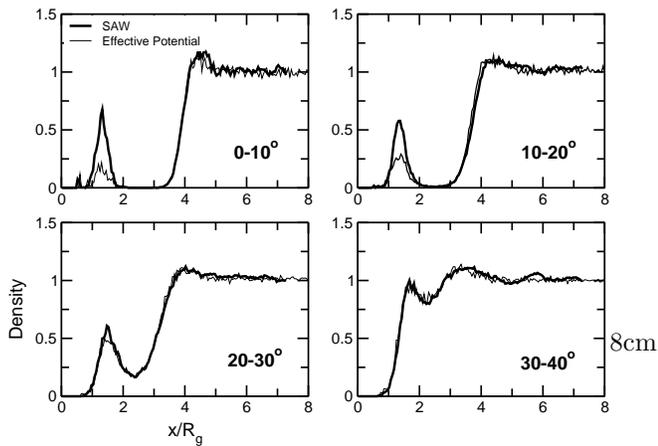}
\caption{\label{fig:dens-plot} Comparison of the direct $L=500$ SAW
simulations of the polymer CM density profile to direct simulations
within the polymers as soft colloids approach for $q=1.05$ and
$\rho/\rho*=2.32$.  One sphere is at $0$, and the other is at $r=2.6
R_g$.  The four graphs show the normalized density at a distance $r$,
averaged over $(0-10\deg)$, $(10-20\deg)$, $(20-30\deg)$, and
$(30-40\deg)$ w.r.t.\ the particle at the origin.  The SAW polymers
penetrate more easily between the colloids than the effective
soft-colloids do.  }
\end{minipage}{8cm}
\end{center}
\end{figure}

The simulation of the depletion potentials induced by the ``soft
colloids'' proceeds in a similar manner to the simulations with SAW
polymers, described in section\ref{SAWdepletion}.  However, the
implementation is much simpler, because the interactions are smooth
and both types of particles are in continuous space.  The simulations are
also about two orders of magnitude faster, since each polymer is
represented by only one particle, instead of the $500$ units per
polymer used for the SAW model.  These simulations are compared for
$q=1.05$ in Fig.~\ref{fig:vrad16-ejm-eff} to the direct SAW
simulations.  The agreement is very good for $\rho/\rho* = 0.58$ and
reasonable for $\rho/\rho* = 1.16$.  However for the highest density,
$\rho/\rho* = 2.32$, the soft-colloid approach  breaks down.
Exactly the same trend was found for the other two size ratios (not
shown here). Since the soft-colloid approach reproduces the correct
one-body densities $\rho(r)$, and the correct one-body free energy
$F_1^{int}(\rho)$ of immersing a single sphere into a bath of polymers
for all densities $\rho/\rho*$ in the dilute and semi-dilute regimes,
it is perhaps surprising that for the case of two spheres, it breaks
down rather abruptly for higher densities.  To further investigate
this issue, we plot the polymer CM density profiles for two spheres
fixed at distance $r= 2.6 R_g$ for $q=1.05$ and $\rho/\rho* = 2.32$ in
Fig.~\ref{fig:dens-plot}. This corresponds to a distance where the SAW
$V(r)$ has effectively gone to zero.  At this short distance the SAW
polymers clearly penetrate more easily in between the two colloids
than the soft particles do, resulting to less attraction between the
colloids. A similar effect was found for polymers between two
walls\cite{Bolh01}, but there the fact that sum-rules constrain the
value of the potential at contact means that this difference
manifested itself instead in an enhanced repulsive bump in the
potential. In both cases, we attribute the error in the soft-colloid
approach at strong confinement and higher densities to a breakdown in
the ``potential superposition approximation''\cite{Bolh01}, i.e.\ the
assumption that a polymer confined between two surfaces feels an
interaction which is simply the sum of the two interactions $\phi(r)$
with each separate surface.  While this would be correct for simple
fluids, it is not correct here, since the close presence of one wall
affects the interaction of the polymers with the second wall. This is
essentially due to the deformability of the polymers, which is not
correctly taken into account for strong confinement.

When we do a similar analysis of the densities as done in
Fig.~\ref{fig:dens-plot} for $\rho/\rho*=1.16$ however, the differences in
the density profiles are very small, suggesting that the soft-particle
picture is not yet breaking down.  Coarse-graining polymers as soft
colloids is expected to be most useful for $q \leq 1$, since once the
polymers become much larger than the colloids, they can easily wrap
around the colloids, an effect not easily treated in our
coarse-graining scheme.  Since for $q \leq 1$, we have found that
phase-separation sets in below $\rho/\rho* \approx 1$\cite{Bolh02},
the limit of physical stability is reached before we encounter
problems with our coarse-graining scheme for polymer-colloid mixtures.

\subsubsection{Depletion potentials from the superposition approximation}

To understand further the nature of the depletion potentials between
the colloidal particles, mediated by the polymers in the ``soft
colloid'' picture, we now turn to the superposition approximation,
which expresses the two body depletion interactions in terms of
one-body properties of a single colloid in a bath of soft polymers.

If one colloidal particle is fixed at the origin, and one is fixed at
{\bf r}, then one can define an inhomogeneous one-particle density of
the soft colloids $\rho^{(1)}({\bf r}';{\bf 0}, {\bf r})$ for that
fixed configuration of the two colloidal particles.  The effective
force induced between two colloidal HS by the soft colloids can then
be written as\cite{Atta89,Bibe96}:
\begin{equation}\label{eq4.1}
{\bf F}({\bf r}) = -\int \rho^{(1)}({\bf r}';{\bf 0}, {\bf
r}) \frac{\partial}{\partial {\bf r}'} \phi(r') d{\bf r}',
\end{equation}
were $\phi(r)$ is the interaction between the HS particles and the
soft colloids.  This expression has an obvious intuitive
interpretation.  The total force on a HS particle at ${\bf 0}$ is the
average of the sum of its interactions with all the soft colloids.  If
there are no other HS particles, then the equilibrium distribution
will be spherically symmetric, and this force will be zero.  However,
if there is another HS particle at ${\bf r}$, it will perturb the
density distribution of the soft colloids, which will in turn result
in a different total force acting on the particle at ${\bf 0}$.  By
definition, this is the effective or depletion force ${\bf F(r)}$
acting on a particle at ${\bf 0}$, induced by the presence of a particle at ${\bf r}$.

\begin{figure}
\begin{center}
\epsfig{figure=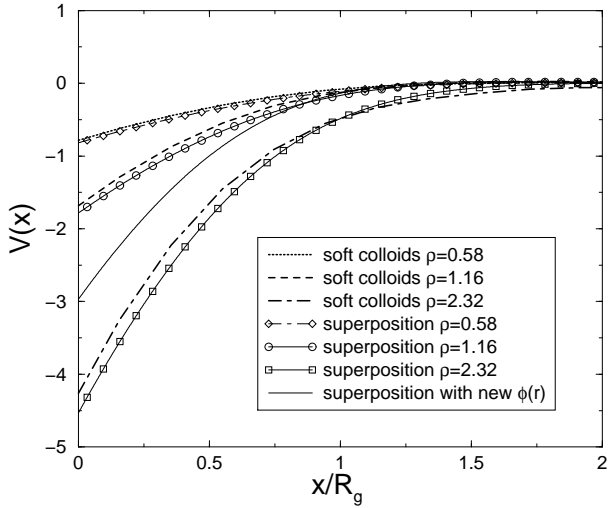,width=8cm}
\caption{\label{fig:vrad16-eff-sup} Comparison of simulations of the
depletion potential for $q=1.05$ within the soft-colloid approach to
the superposition approximation of Eqs.~(\protect\ref{eq4.1})
and~\protect\ref{eq4.2}).  The thin solid line denotes the
superposition calculation with $\rho(r)$ from $\rho/\rho*=2.32$, but
the potential $\phi(r)$ from the $\rho/\rho*=1.16$ calculation.  This
illustrates the sensitivity of the depletion potentials to
colloid-polymer CM interaction $\phi(r)$.
}
\end{center}
\end{figure}

 Eq.~(\ref{eq4.1}) is in principle exact, but requires knowledge of
the soft colloid density around two  colloidal HS.  By
approximating the full (cylindrically symmetric) one-body density
$\rho^{(1)}({\bf r}';{\bf 0}, {\bf r})$ by a superposition of the
one-body density $\rho(r)$ around an isolated single
sphere\cite{Atta89}:
\begin{equation}\label{eq4.2}
\rho^{(1)}({\bf r}';{\bf 0}, {\bf r}) = \rho( r') \rho(|{\bf r -
r}'|)/\rho,
\end{equation}
the effective depletion force (\ref{eq4.1}) can be entirely expressed
in terms of the (radially symmetric) problem of a single colloidal
sphere immersed in a polymer solution. The results of such a
calculation for $V(r)$ are compared in Fig.~\ref{fig:vrad16-eff-sup}
for $q=1.05$.  For all three densities, the superposition
approximation closely follows the direct simulations for the
soft-colloids.  This implies that the full one-body density
$\rho^{(1)}({\bf r}';{\bf 0}, {\bf r})$ does not differ much from the
simple superposition of Eq.~(\ref{eq4.2}), even for the highest
density considered, $\rho/\rho*=2.32$.  The reason for this is most
likely that the effective  polymer--polymer CM interaction $v(r)$ is
rather weak.  In a recent study, a similar good performance for the
superposition approximation was found for a low density hard-core
depletant when the HS-depletant interaction was fairly long
ranged\cite{Loui01f}.  Good agreement was also found in a similar study
of star-polymer colloid mixtures\cite{Dzub01} where accurate
star-polymer--wall potentials\cite{Jusu01} are derived based on a
coarse-graining approach\cite{Liko98}.  In contrast, if the density of
a hard-core depletant increases, then the superposition approximation
is known to break down for increasing packing
fraction\cite{Bibe96,Amok98}.

The fact that the superposition approximation works well for the soft
colloids at the higher densities, also raises an interesting question
as to why the soft-colloid picture itself breaks down at the higher
densities.  By construction, the one-body density profiles are
identical for the full polymer case and the soft-colloid
picture. Presumably the reason for the difference in the two-body
depletion interactions can be traced either to a large deviation from
superposition for the full polymer case, or else to a breakdown of the
potential superposition approximation, as discussed in
ref.~\cite{Bolh01}. The influence of changing $\phi(r)$ is illustrated
in Fig.~\ref{fig:vrad16-eff-sup}, where we used the $\phi(r)$
appropriate for $\rho/\rho*=1.16$ to calculate the depletion potential
with the density profiles for $\rho/\rho*=2.32$.  This $\phi(r)$ is
less repulsive than the true $\phi(r)$, and leads to an important
difference in the depletion potential.  These depletion interactions
are therefore quite sensitive to the exact form of $\phi(r)$.

In conclusion then, the superposition approximation works remarkably
well  for $\rho/\rho* \leq 1$, the regime where the
soft-colloid picture provides the most accurate representation of the
true depletion potentials.  Since the density profiles $\rho(r)$ can
be easily calculated by our wall-OZ approach, this means that one only
needs the polymer-colloid interaction $\phi(r)$ and the
polymer-polymer interaction $v(r)$ as input to reliably calculate the
full two-body depletion interactions from the superposition
approximation.

\section{Second virial coefficients  phase-diagrams}

\begin{figure}
\begin{center}
\epsfig{figure=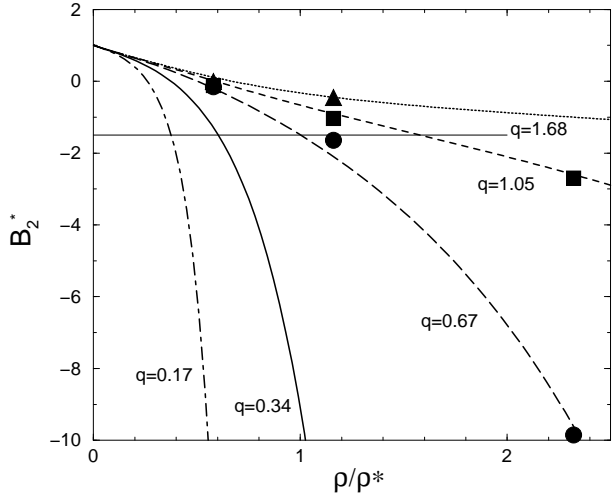,width=8cm}
\caption{\label{fig:B2} Virial coefficients: the symbols represent
results from the computer simulations of L=500 SAW polymers for
$q=0.67$ (circles), $q=1.05$ (squares), and $q=1.68$ (triangles). The
straight lines are from Eq.~(\protect\ref{eq6.2}).  The good agreement
shows that Eq.~(\protect\ref{eq3.21}) provides a good
representation of the depletion potential. }
\end{center}
\end{figure}

A useful measure of the strength of an interaction is the second
osmotic virial coefficient $B_2$, since this is experimentally
accessible through a measurement of the osmotic pressure at low
densities\cite{deHek82,Kulk99}.  Such virial coefficients are very
sensitive to the strength and nature of the depletion
interaction\cite{Roth01}.  A recent study\cite{Vliegenthart00} has
shown that for many simple systems consisting of a HS like repulsion
with an additional attractive potential, the location of the
liquid-gas or fluid-fluid critical point can be correlated with the
point where the reduced virial coefficient $B_2^* = B_2/B_2^{HS}
\approx -1.5$, where $B_2^{HS} = 16 \pi R_c^3 /3$ is the virial
coefficient of a HS system with radius $R_c$.  For our depletion
systems, the reduced virial coefficient is given by
\begin{equation}\label{eq6.1}
B_2^* = 1 - \frac{3}{4 R_c^3} \int_{R_c}^{\infty} x^2
\left(\exp[-V(x)] -1\right).
\end{equation}
The direct simulation results depicted in
Fig~\ref{fig:vrad16-ejm-newpot} can be used to calculate $B_2^*$, and
the results are shown in Fig~\ref{fig:B2} for $\rho/\rho*=0.58$,
$\rho/\rho*=1.16$, and for  $\rho/\rho*=2.32$.   As expected, for a
given density $\rho/\rho*$, the $B_2^*$ become progressively more
negative with decreasing $q$.  Although we don't include any explicit
error bars in our plots, these may be rather large for increasing
attraction because they appear exponentially in Eq.~(\ref{eq6.1}).

We found in the previous sections that a simple semi-empirical
potential $V_s(x)$, given by Eq.~(\ref{eq3.21}), provided a nearly
quantitative description of the depletion potentials between two
spheres.  It turns out that for this potential the virial coefficient
can be integrated analytically:
\begin{eqnarray}\label{eq6.2}
\frac{B_2}{B_2^{HS}} &= & \frac{1}{16\pi R_c^{9/2}W^{3/2}}
\left( 2 \sqrt{R_cW} \left( -6 D (D + 2R_c) \right. \right.\nonumber
\\ &+ & \left. 3D \exp\left[ \frac{\pi}{2} R_c W D\right] (D + 4 R_c)
+ \pi R_c W (D + 2R_c)^3 R_c \right) \nonumber \\ 
&+& \left. 3 \sqrt{2 D} (D -\pi R_c W (D + 2 R_c)^2)
Erfi\left[\sqrt{\frac{\pi R_c D W}{2}} \right] \right)
\end{eqnarray}
where $W=W(x=0)$ and $D=D_s(\rho)$, i.e. the explicit density
dependence has been suppressed for notational clarity. The imaginary
error function $Erfi[z]$ is defined as $Erf[iz]/i$;  its value is
real. As can be seen in Fig.~\ref{fig:B2}, this expression compares
well with the direct simulation results, demonstrating that the
potential~(\ref{eq3.21}) is a good approximation to the true depletion
potential.  For this reason, we also plot the theoretical $B_2$ for
two smaller $q$'s. The density $\rho/\rho*$ at which one would expect
a (metastable) critical point and a related fluid-fluid de-mixing
decreases with decreasing $q$.  To compare with experiments, one
should keep in mind that our parameter $\rho/\rho*$ would be equal to
the density of a reservoir of polymers kept at the same chemical
potential as a full polymer-colloid mixture\cite{murho}.


\vspace*{-0.5cm}

If the pair-potentials are of sufficiently short range, the
fluid-fluid transition becomes metastable w.r.t.\ the fluid-solid
transition of the larger particles\cite{Gast83,Lekk92,Hage94}, which
widens to a much larger.  It has been recently found that this generic
widening of the fluid-solid transition is primarily driven by the
value of the potential at contact\cite{Loui00a,Loui01a}; other details
such as the shape and range of the attractive part of the potential
are less important.  Roughly speaking, the fluid-solid liquidus curve
widens to about half the packing fraction of the hard-sphere freezing
transition when $V(0) \approx 2.5 \pm 0.5$\cite{Loui01a}.  When
combined with the criterion that $B_2^* \approx -1.5$ at the critical
density, this can be used for a rough prediction of the $q$ below
which the fluid-fluid transition is metastable w.r.t.\ the fluid-solid
one.  The minimum $q$ for which the fluid-fluid transition is still
stable is between $0.3$ and $0.5$, which is consistent with other
calculations\cite{Gast83,Lekk92,Meij94,Dijk99}, and
experiments\cite{Cald93,Ilet95}.  Of course one should keep in mind
that the larger the size ratio $q$, the more important three-body and
higher order interactions become\cite{Meij94,Goul01b}.  These
pair-potential criteria should therefore become less accurate for
increasing $q$.

\begin{figure}
\begin{center}
\epsfig{figure=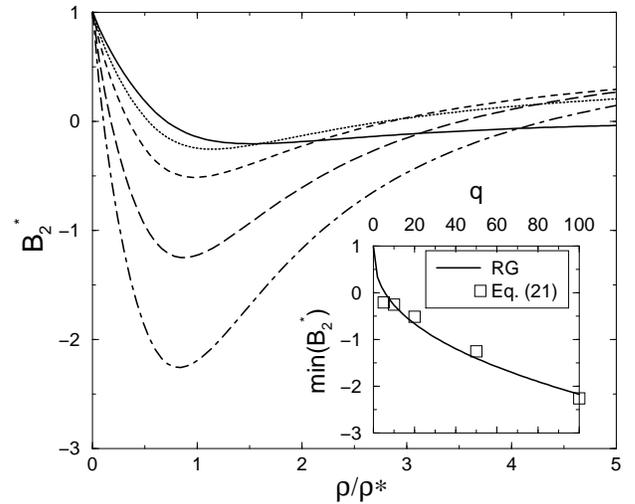,width=8cm}
\caption{\label{fig:B2-smallq} Virial coefficient from
Eq.~(\protect\ref{eq6.2}) as a function of polymer density for large
size ratios: $q=5$ (solid line), $q=10$ (dotted line) $q=20$ (dashed
line), $q=50$ (long-dashed line) and $q=100$ (dot-dashed line).
Inset: Comparison of the minimum of $B_2^*$ as a function of $q$ to
the asymptotic RG result $min(B_2^*) = 1 - 0.5q^{0.401}$.  The agreement is
remarkable.  }
\end{center}
\end{figure}

 Both RG theory\cite{Eise00} and integral equation
approaches\cite{Chat98,Fuch01} predict that for large enough $q$,
$B_2^*$ should go through a minimum as a function of $\rho/\rho*$,
although the two theories differ significantly on quantitative
details.  Experiments seem to show similar effects\cite{Kulk99}.
Remarkably, expression~(\ref{eq6.2}), which is based on the simple
potential of Eq.~(\ref{eq3.21}) also shows a minimum, as shown in
Fig.~\ref{fig:B2-smallq}.  Even more surprisingly, considering how the
potential was derived, is that the minimum is near $\rho/\rho*=1$, as
predicted by RG theory, and that it closely follows the asymptotic law
$min(B_2^*) = 1 - 0.5 q^{0.401}$, derived by Eisenriegler\cite{Eise00}
(see the inset of Fig.~\ref{fig:B2-smallq}).  In contrast, when the
simpler Derjaguin expression~(\ref{eq3.10}) is used for $B_2^*$, no
minimum as a function of $\rho/\rho*$ appears for large $q$.  At
present it is not clear why our expression for $B_2^*$,
Eq.~(\ref{eq6.2}), should agree so well with the RG results in this
large $R_g/R_c$ regime.

\section{Discussion and Conclusion}

We have used a combination of computer simulations and theory to
derive the depletion interactions induced by excluded volume polymers.
Our main computer simulation results are depicted in
Fig.~\ref{fig:Forces} and Fig.~\ref{fig:vrad16-ejm-newpot} and our
main theoretical results are the depletion potential for two plates,
given by Eq.~(\ref{eq2.6}) and the depletion potential for two
spheres, given by Eq.~(\ref{eq3.21}).

For two plates, we found that a simple linear depletion
potential~(\ref{eq2.6}) with a well-depth of $W(0) = 2
\gamma_w(\rho)$, and a slope of $\Pi(\rho)$ is very accurate.  The
range decreases with density, with the largest relative change occurring in the
dilute regime.  In the semi-dilute regime, the well-depth scales as
$W(0) \sim \rho^{1.539}$, while the range scales as $D_w(\rho) \sim
\rho^{-0.770}$.  The depletion potential also takes on a simpler
functional form, given by Eq.~(\ref{eq2.7}). These results differ
significantly from theories based on ideal polymers.

For two spheres, we find that the simple Derjaguin approximation works
much better than expected due to a cancellation of errors related to
the deformation of the polymers around a sphere.  Interestingly, the
Derjaguin expression for the force at contact seems particularly
accurate, even for very large size ratios $q$.  We used this
observation, together with a correction for the Derjaguin range, to
derive the depletion potential $V_s(r)$ (Eq.~(\ref{eq3.21})), which
was shown to agree quantitatively with our direct computer simulation
results.  While the excellent agreement with simulations is
gratifying, further study is needed to understand why our main ansatz,
that the force at contact is accurately approximated by the Derjaguin
approximation, works so well for such a wide range of $q$.

  The range of our potential scales with density in the same way that
the range for the two plates does. But in contrast to the case of two
plates, where the well depth at contact is more attractive than that
for ideal polymers, the well depth at contact for two spheres is less
attractive than that predicted by ideal polymer theories.  For
example, in the semi-dilute regime, $V_s(0) \sim -\rho^{0.770}$,
compared to the scaling of the AO potential: $V_{AO}(0) \sim -\rho$.
In other words, ideal polymer theories overestimate both the range
{\em and} the well-depth of the depletion potential between two
spheres, induced by interacting polymers.

We also calculated the depletion potential between spheres induced by
polymers coarse-grained as ``soft colloids''.  For the dilute regime,
good agreement with the direct simulations of SAW polymers was
achieved.  However, for the highest density, $\rho/\rho* = 2.32$, the
polymers as soft colloids approach strongly overestimates the depth of
the depletion potential.  We traced this discrepancy to a breakdown of
the potential superposition approximation, something also found for
two plates\cite{Bolh01}.  However, since our coarse-graining scheme is
mainly useful for $q \leq 1$, where phase-separation sets in for
$\rho/\rho* \leq 1$, this breakdown at high density is not relevant to
the equilibrium phase behavior of colloid-polymer systems.  In fact,
we have recently used this scheme to calculate the phase-separation
binodals of polymer-colloid mixtures for several size ratio's $q \leq
1$\cite{Bolh02}.  Direct simulations of SAW polymers and colloids
would be significantly more expensive. The advantage over directly using a
pair-potential such as Eq.~(\ref{eq3.21}), is that many-body
interactions are also effectively taken into account\cite{Bolh02}.

Virial coefficients can provide a sensitive measure of the quality of an
effective potential.  We were able to integrate Eq.~(\ref{eq3.21}) to
derive an analytical representation of the virial
coefficient~(\ref{eq6.2}), which compares very well with the direct
simulation results.  Surprisingly, this same analytic form agrees
quantitatively with RG results in the limit $q \gg 1$.  Further study
is needed to clarify why Eq.~(\ref{eq6.2}) works so well in this
limit.

A careful comparison with experiments is complicated by the fact that
most measurements are of phase behavior or structure, which  depend only
indirectly on the depletion interactions.  The most extensive direct
measurements of the depletion potential between spheres, carried out
by Verma {\em et al.}\cite{Verm98}, were on a system with
semi-flexible polymers (DNA) which do not follow the same scaling laws as
the SAW polymers we treated in this paper.  However, given a good
model for the equation of state and the adsorption $\hat{\Gamma}(\rho)$
for these polymers, it would be straightforward to calculate the
depletion potentials using methods derived in this paper.  A detailed
comparison with experiments will be the subject of an upcoming
review\cite{Loui02}.

This paper has concerned itself with the case of interacting,
non-adsorbing polymers in a good solvent.  By taking
Eqs.~(\ref{eq2.1}), (\ref{eq2.2}), (\ref{eq2.6}), and (\ref{eq3.21})
together, if follows that accurate depletion potentials can be
calculated with only a knowledge of the adsorption
$\hat{\Gamma}(\rho)$ and the equation of state $\Pi(\rho)$.  This
implies that the effect of changing solvent quality, or of adding
attractions or repulsions between the colloids and the polymers can be
understood by seeing how they effect these two quantities.  Note that
this is different from binary colloid mixtures, where the effects of
added interparticle attractions and repulsions have a more subtle
effect on the resultant depletion potentials\cite{Loui01f}. 

 It is not hard to see that for polymers in a good solvent, adding an
attractive (repulsive) polymer-colloid interaction results in a less
(more) attractive depletion potential, and possibly even to repulsive
effective interactions.  This follows since $\Pi(\rho)$ is unchanged,
and only $\hat{\Gamma}(\rho)$ is affected in an obvious way. An
example where this may be relevant concerns the common practice to
stabilize colloidal particles by a short polymer brush.  The
interaction of the free polymers with the polymer brush can be quite
subtle\cite{Jone89}, but our theory implies that it is only the
overall change in the adsorption which effects the depletion
potential, and by extension, the phase behavior.

 Changing the solvent quality has a less obvious effect, since both
$\Pi(\rho)$ and $\hat{\Gamma}\rho)$ are affected. For a given
density, we expect the absolute magnitudes of both to decrease, so
that the strength of the depletion potential should decrease as well.
Of course, for strong deviations from the case of non-adsorbing
polymers in a good solvent, some of the simplifying assumptions that
went in to the derivation of the depletion potentials may begin to
break down.  For example, for very strong adsorption, the system may
show bridging flocculation\cite{Brou00}, or gelation\cite{John01}. A
rich phenomenology may be expected; this will be the subject of an
upcoming investigation\cite{Krak02}.

\section*{Acknowledgements} 
AAL acknowledges support from the Isaac Newton Trust, Cambridge, and
the hospitality of Lyd\'{e}ric Bocquet at the Ecole Normale Superieure
in Lyon, where much of this work was carried out.  EJM acknowledges
support from the Royal Netherlands Academy of Arts and Sciences.  and
support from the Stichting Nationale Computerfaciliteiten (NCF) and
Nederlandse Organisatie voor Wetenchappelijk Onderzoek (NWO) for the
use of supercomputer facilities.  We thank R. Evans and H. L\"{o}wen
for helpful discussions.

\appendix
\section{Derjaguin approximation for the Asakura-Oosawa model}
\subsection{Comparison for fixed parameter $R_{AO}$}

The Asakura Oosawa model, where the ideal polymers are modeled as
interpenetrable spheres of radius $R_{AO}$, was first introduced
in 1958\cite{Asak58}.  Between two plates this results in the linear
depletion potential similar to Eq.~(\ref{eq2.10}), with a range
$R_{AO}$, while between two spheres it results in the well-known
Asakura-Oosawa form
\begin{eqnarray}\label{eqA.1}
V_{AO}(x) & = & - \rho\frac{4 \pi}{3}\left(\sigma_{cp}\right)^3
\left\{ 1 - \frac{3}{4} \left(\frac{x + 2 R_c}{\sigma_{cp}}\right)
\right. \nonumber \\ &+ & \left. \frac{1}{16} \left(\frac{x + 2
R_c}{\sigma_{cp}}\right)^3 \right\} \,\,\,\,\,\,\, 0 \leq x \leq 2 R_{AO}
\end{eqnarray}
which is exact within the confines of the AO model. Here we defined
$\sigma_{cp} = R_c + R_{AO}$.  Similarly, the use of  the Derjaguin
approximation (\ref{eq3.2}) together with the linear AO potential between two
plates results in:
\begin{equation}\label{A.2}
V_{AO}^{Derj}(x) = -\rho \frac{\pi}{2} R_c \left(2 R_{AO}-x\right)^2
\,\,\,\,\,\,\, 0 \leq x \leq 2 R_{AO}.
\end{equation}
 Comparison of the two expressions at contact, where $V_{AO}(0)= -
\rho 2 \pi (R_c R_{AO}^2 + 2 R_{AO}^3/3)$ and $V_{AO}^{Derj}(0)= -\rho
2 \pi R_c R_{AO}^2$, shows that the Derjaguin expression (\ref{A.2})
underestimates the well depth at contact w.r.t.\ the exact
expression(\ref{eqA.1}), a discrepancy which gets relatively worse for
decreasing $R_c/R_{AO}$.  This is illustrated in
Fig~\ref{fig:VAO-Derj}(a) for the same size ratios we used for the
interacting polymers.  Since the Derjaguin approximation is only valid
for large $R_c/R_{AO}$, this breakdown is expected.

\subsection{Comparison for effective parameter $R_{AO}^{eff}(R_g,R_c)$}

Before using the AO model to describe polymers, one needs to fix the
parameter $R_{AO}$.  A common prescription has been to take
$R_{AO}=R_g$\cite{Asak58}.  However, between two walls, one should
take $R_{AO} = (2/\sqrt{\pi}) R_g$, which corresponds to approximating
the depletion layer of ideal polymers by a step-function with the same
depletion volume. This then results in a well-depth which is equal to
that found for ideal polymers\cite{Bolh01}.

However, for polymers of a given size $R_g$, the width of the
depletion layer around a single sphere will decrease with decreasing
$R_c$, because the polymers can partially wrap around the colloidal
particles\cite{Meij94,Eise96}.  By equating the one-body insertion
free energy for a HS sphere into a bath of AO model particles to that
found for inserting a HS into a bath of ideal polymers, we found, in
paper I, an exact analytic prescription for the effective parameter
$R_{AO}^{eff}$
\begin{equation}\label{eqA.4}
\frac{R_{AO}^{eff}}{R_g} = \frac{1}{q} \left( \left(1 +
 \frac{6}{\sqrt{\pi}} q + 3 q^2 \right)^{1/3}
 -1 \right),
\end{equation}
which decreases with decreasing $R_c$ as expected.  Just because the
one-body terms are equal does not mean that the same prescription
for $R_{AO}$ will work for the two-body terms.  But direct computer
simulations\cite{Meij94} have shown that using $R_{AO}^{eff}$ in the
full AO expression (\ref{eqA.1}) results in a much better
approximation of the depletion potential between two spheres in a bath
of ideal polymers, than using a prescription which fixes $R_{AO}$
independently of $R_c$.

\begin{figure}
\begin{center}
\epsfig{figure=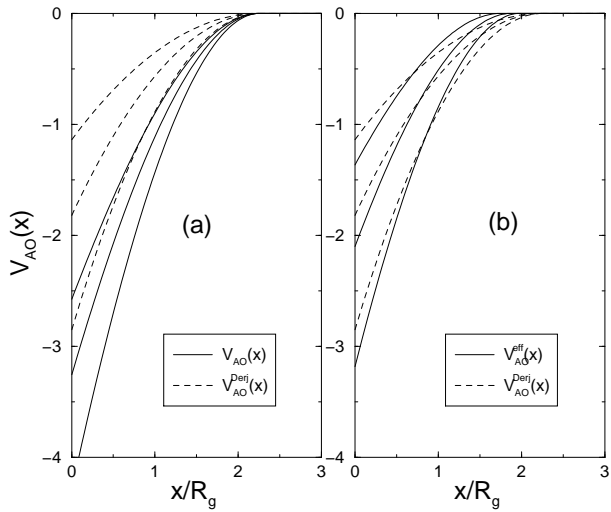,width=8cm}
\caption{\label{fig:VAO-Derj} {\bf (a)} Comparison of the full
Asakura-Oosawa model interaction (\protect\ref{eqA.1}) with $R_{AO} =
2/\sqrt{\pi}$ to the Derjaguin approximation expression
(\protect\ref{A.2}) with the same $R_{AO}$.  The hard sphere sizes are
the same as in Fig.~\protect\ref{fig:vrad16-ejm-newpot}. The
potentials, from bottom to top, correspond to $q=1.68$, $q=1.05$ and
$q=0.67$ respectively. The polymer density is $\rho/\rho*=1$.  {\bf
(b)} The same comparison as in {\bf (a)}, but now the decrease of the
depletion layer range with decreasing sphere size is taken into
account by using the effective AO parameter $R_{AO}^{eff}$, defined in
Eq.~(\protect\ref{eqA.4}), in the full AO
expression~(\protect\ref{eqA.1}).  For the Derjaguin approximation
expression $R_{AO} =2/\sqrt{\pi}$ is unchanged.  The agreement is now
much better.  }
\end{center}
\end{figure}

With this in mind, we now compare again the Derjaguin expression
(\ref{A.2}) with $R_{AO} = 2/\sqrt{\pi} R_g$ to the full AO expression
(\ref{eqA.1}) but now with an effective AO radius $R_{AO}^{eff}$ from
Eq.~(\ref{eqA.4}) appropriate to the desired $R_c$.  As demonstrated
in Fig.~\ref{fig:VAO-Derj}(b), the Derjaguin approach is now a more
faithful approximation of the full AO model depletion potential, which
is itself a better approximation of the true depletion potential
induced by ideal polymers.  Although the Derjaguin approach doesn't
capture the slight reduction in range, the well-depth is much closer
to the  AO value.

Of course this improvement arises from a cancellation of errors.
However, for the interacting polymers the depletion potential between
two walls is also well described by a linear form, and we expect the
same deformation effect to be relevant for polymers near
spheres.  Therefore, we might hope for a similar
cancellation of errors when the Derjaguin expression is used for
interacting polymers.

\end{multicols}

\begin{thebibliography}{99}
\bibitem{murho} It is generally most natural to analyze depletion
systems in a semi-grand ensemble, where the colloidal particles are in
osmotic equilibrium with a reservoir of polymers at fixed chemical
potential $\mu$\protect\cite{Lekk92,Dijk98}. Throughout this paper,
therefore, the polymer density $\rho$ is actually that of a reservoir
of polymers at the same chemical potential $\mu$.  For the
two-particle depletion problem in the thermodynamic limit, this is
equivalent to the actual polymer density, but if one wants to use the
depletion potential to analyze the phase behavior of a many colloid
system, this prescription should be kept in mind.
\bibitem{zero} It would be straightforward to generalize the potential
to one with a slightly more rounded shape, such as that found for ideal
polymers\protect\cite{Asak54,Bolh01}.  Although this would be  slightly more
accurate, the differences are so small that we will stick to the
simpler linear form for this paper.  The definition for the range 
would remain the same for both versions of the potential.
\bibitem{Dickcompare} To compare to the results
of~\protect\cite{Dick94} we used their expression for the end-to-end
distance $R_e^0$ for the three dimensional fluctuating bond model, to
derive the radius of gyration at zero density through $R_g =
R_e^0/\sqrt{6.06}$\protect\cite{Doi86}.  The two results with the
highest monomer density are not shown since they appear to be closer
to the melt than to the semi-dilute regime.
\end{thebibliography}
\end{document}